
\magnification=\magstephalf
\baselineskip=13.5pt

\hsize=6.5 truein
\vsize=9.5 truein
\hfuzz=2pt\vfuzz=4pt
\pretolerance=5000
\tolerance=5000
\parskip=0pt plus 1pt
\parindent=16pt
\font\fourteenrm=cmr10 scaled \magstep2
\font\fourteeni=cmmi10 scaled \magstep2
\font\fourteenbf=cmbx10 scaled \magstep2
\font\fourteenit=cmti10 scaled \magstep2
\font\fourteensy=cmsy10 scaled \magstep2
\font\large=cmbx10 scaled \magstep1

\font\eightrm=cmr8
\font\eighti=cmmi8
\font\eightbf=cmbx8
\font\eightit=cmti8

\font\eightsy=cmsy8

\font\sixrm=cmr6
\font\sixi=cmmi6
\font\sixsy=cmsy6

\def\tenpoint{\def\rm{\fam0\tenrm}%
  \textfont0=\tenrm \scriptfont0=\sevenrm
		      \scriptscriptfont0=\fiverm
  \textfont1=\teni  \scriptfont1=\seveni
		      \scriptscriptfont1=\fivei
  \textfont2=\tensy \scriptfont2=\sevensy
		      \scriptscriptfont2=\fivesy
  \textfont3=\tenex   \scriptfont3=\tenex
		      \scriptscriptfont3=\tenex
  \textfont\itfam=\tenit  \def\it{\fam\itfam\tenit}%
  \textfont\slfam=\tensl  \def\sl{\fam\slfam\tensl}%
  \textfont\bffam=\tenbf  \scriptfont\bffam=\sevenbf
			    \scriptscriptfont\bffam=\fivebf
			    \def\bf{\fam\bffam\tenbf}%
  \normalbaselineskip=20 truept
  \setbox\strutbox=\hbox{\vrule height14pt depth6pt width0pt}%
  \let\sc=\eightrm \normalbaselines\rm}
\def\eightpoint{\def\rm{\fam0\eightrm}%
  \textfont0=\eightrm \scriptfont0=\sixrm
		      \scriptscriptfont0=\fiverm
  \textfont1=\eighti  \scriptfont1=\sixi
		      \scriptscriptfont1=\fivei
  \textfont2=\eightsy \scriptfont2=\sixsy
		      \scriptscriptfont2=\fivesy
  \textfont3=\tenex   \scriptfont3=\tenex
		      \scriptscriptfont3=\tenex
  \textfont\itfam=\eightit  \def\it{\fam\itfam\eightit}%
  \textfont\bffam=\eightbf  \def\bf{\fam\bffam\eightbf}%
  \normalbaselineskip=16 truept
  \setbox\strutbox=\hbox{\vrule height11pt depth5pt width0pt}}
\def\fourteenpoint{\def\rm{\fam0\fourteenrm}%
  \textfont0=\fourteenrm \scriptfont0=\tenrm
		      \scriptscriptfont0=\eightrm
  \textfont1=\fourteeni  \scriptfont1=\teni
		      \scriptscriptfont1=\eighti
  \textfont2=\fourteensy \scriptfont2=\tensy
		      \scriptscriptfont2=\eightsy
  \textfont3=\tenex   \scriptfont3=\tenex
		      \scriptscriptfont3=\tenex
  \textfont\itfam=\fourteenit  \def\it{\fam\itfam\fourteenit}%
  \textfont\bffam=\fourteenbf  \scriptfont\bffam=\tenbf
			     \scriptscriptfont\bffam=\eightbf
			     \def\bf{\fam\bffam\fourteenbf}%
  \normalbaselineskip=24 truept
  \setbox\strutbox=\hbox{\vrule height17pt depth7pt width0pt}%
  \let\sc=\tenrm \normalbaselines\rm}

\def\today{\number\day\ \ifcase\month\or
  January\or February\or March\or April\or May\or June\or
  July\or August\or September\or October\or November\or December\fi
  \space \number\year}
\newcount\secno      
\newcount\subno      
\newcount\subsubno   
\newcount\appno      
\newcount\tableno    
\newcount\figureno   
\normalbaselineskip=15 truept
\baselineskip=15 truept
\def\title#1
   {\vglue1truein
   {\baselineskip=24 truept
    \pretolerance=10000
    \raggedright
    \noindent \fourteenpoint\bf #1\par}
    \vskip1truein minus36pt}
\def\author#1
  {{\pretolerance=10000
    \raggedright
    \noindent {\large #1}\par}}
\def\address#1
   {\bigskip
    \noindent \rm #1\par}
\def\shorttitle#1
   {\vfill
    \noindent \rm Short title: {\sl #1}\par
    \medskip}
\def\pacs#1
   {\noindent \rm PACS number(s): #1\par
    \medskip}
\def\jnl#1
   {\noindent \rm Submitted to: {\sl #1}\par
    \medskip}
\def\date
   {\noindent Date: \today\par
    \medskip}
\def\beginabstract
   {\vskip .5in\baselineskip=12pt
    \noindent {\bf Abstract. }\smallskip\rm}
\def\keyword#1
   {\bigskip
    \noindent {\bf Keyword abstract: }\rm#1}
\def\endabstract
   {\par
    \vfill\eject}

\def\entry#1#2#3
   {\noindent
    \hangindent=20pt
    \hangafter=1
    \hbox to20pt{#1 \hss}#2\hfill #3\par}
\def\subentry#1#2#3
   {\noindent
    \hangindent=40pt
    \hangafter=1
    \hskip20pt\hbox to20pt{#1 \hss}#2\hfill #3\par}
\def\section#1
   {\vskip0pt plus.1\vsize\penalty-250
    \vskip0pt plus-.1\vsize\vskip24pt plus12pt minus6pt
    \subno=0 \subsubno=0
    \global\advance\secno by 1
    \noindent {\bf \the\secno. #1\par}
    \bigskip
    \noindent}
\def\subsection#1
   {\vskip-\lastskip
    \vskip24pt plus12pt minus6pt
    \bigbreak
    \global\advance\subno by 1
    \subsubno=0
    \noindent {\sl \the\secno.\the\subno. #1\par}
    \nobreak
    \medskip
    \noindent}
\def\subsubsection#1
   {\vskip-\lastskip
    \vskip20pt plus6pt minus6pt
    \bigbreak
    \global\advance\subsubno by 1
    \noindent {\sl \the\secno.\the\subno.\the\subsubno. #1}\null. }
\def\appendix#1
   {\vskip0pt plus.1\vsize\penalty-250
    \vskip0pt plus-.1\vsize\vskip24pt plus12pt minus6pt
    \subno=0 \eqnno=0
    \global\advance\appno by 1
    \noindent {\bf Appendix \the\appno. #1\par}
    \bigskip
    \noindent}
\def\subappendix#1
   {\vskip-\lastskip
    \vskip36pt plus12pt minus12pt
    \bigbreak
    \global\advance\subno by 1
    \noindent {\sl \the\appno.\the\subno. #1\par}
    \nobreak
    \medskip
    \noindent}
\def\ack
   {\vskip-\lastskip
    \vskip36pt plus12pt minus12pt
    \bigbreak
    \noindent{\bf Acknowledgements\par}
    \nobreak
    \bigskip
    \noindent}

\def\tabcaption#1
   {\global\advance\tableno by 1
    \noindent {\bf Table \the\tableno.} \rm#1\par
    \bigskip}
\def\figures
   {\vskip 1in 
    \noindent {\bf Figure captions\par}
    \bigskip}
\def\figcaption#1
   {\global\advance\figureno by 1
    \noindent {\bf Figure \the\figureno.} \rm#1\par
    \bigskip}
\def\references
     {\vskip .5in 
     {\noindent \bf References\par}
      \parindent=0pt
      \bigskip}
\def\refjl#1#2#3#4
   {\hangindent=16pt
    \hangafter=1
    \rm #1
   {\frenchspacing\sl #2
    \bf #3}
    #4\par}
\def\refbk#1#2#3
   {\hangindent=16pt
    \hangafter=1
    \rm #1
   {\frenchspacing\sl #2}
    #3\par}
\def\numrefjl#1#2#3#4#5
   {\parindent=40pt
    \hang
    \noindent
    \rm {\hbox to 30truept{\hss #1\quad}}#2
   {\frenchspacing\sl #3\/
    \bf #4}
    #5\par\parindent=16pt}
\def\numrefbk#1#2#3#4
   {\parindent=40pt
    \hang
    \noindent
    \rm {\hbox to 30truept{\hss #1\quad}}#2
   {\frenchspacing\sl #3\/}
    #4\par\parindent=16pt}

\def\frac#1#2{{#1 \over #2}}

\def\d{{\rm d}}

\def\i{\ifmmode{\rm i}\else\char"10\fi}

\def\boldrule#1{\vbox{\hrule height1pt width#1}}

\def\etal{{\sl et al\/}\ }
\catcode`\@=11
\def\ind{\hbox to 5pc{}}
\def\eq(#1){\hfill\llap{(#1)}}

\def\deqn#1{\displ@y\halign{\hbox to \displaywidth
    {$\@lign\displaystyle##\hfil$}\crcr #1\crcr}}
\def\indeqn#1{\displ@y\halign{\hbox to \displaywidth
    {$\ind\@lign\displaystyle##\hfil$}\crcr #1\crcr}}
\def\indalign#1{\displ@y \tabskip=0pt
  \halign to\displaywidth{\ind$\@lign\displaystyle{##}$\tabskip=0pt
    &$\@lign\displaystyle{{}##}$\hfill\tabskip=\centering
    &\llap{$\@lign##$}\tabskip=0pt\crcr
    #1\crcr}}
\catcode`\@=12

\def\IP{Inverse Problems}
\def\JPA{J. Phys. A: Math. Gen.}


\def\JMP{J. Math. Phys.}

\def\JPSJ{J. Phys. Soc. Japan}

\def\PRL{Phys. Rev. Lett.}


\font\sc=cmcsc10

\font\nit=cmti9
\font\nrm=cmr9
\font\nbf=cmbx9
\font\nsl=cmsl9

\font\shell=msbm10
\def\Re{{\hbox{\shell R}}}

\def\p{Pain\-lev\'e}
\def\d{{\rm d}}\def\i{{\rm i}}

\def\tfr#1#2{{\tx{#1\over#2}}}

\newcount\refno
\def\ref#1#2#3#4#5{\vskip.9pt\global\advance\refno by 1
\item{[{\bf\the\refno}]\ }{\rm#1}, {\it#2}, {\bf#3} (#4) #5}

\def\sam{Stud.\ Appl.\ Math.}
\def\pl{Phys.\ Lett.}
\def\jpa{J.\ Phys.\ A: Math.\ Gen.}
\def\jmp{J.\ Math.\ Phys}

\def\~#1{{\bf\tilde{\mit#1}}}


\def\secn{\the\secno}

\font\sit=cmti9

\def\pd#1{{\partial_{#1}}}
\nopagenumbers
\def\sch{Schr\"odinger}

\def\tfr#1#2{{\tx{#1\over#2}}}
\def\pde{PDE}
\def\pdes{PDEs}
\def\ode{ODE}
\def\odes{ODEs}

\def\eq{equa\-tion}

\def\dgb{Differ\-ential Gr\"obner Bases}
\def\d{{\rm d}}\def\i{{\rm i}}

\def\sech{\mathop{\rm sech}\nolimits}

\font\teneuf=eufm10 \font\seveneuf=eufm7 \font\fiveeuf=eufm5
\newfam\euffam
\textfont\euffam=\teneuf \scriptfont\euffam=\seveneuf
\scriptscriptfont\euffam=\fiveeuf
\def\frak#1{{\fam\euffam\relax#1}}

 \def\fg{\frak g} 

\newcount\exno
\newcount\secno   
\newcount\subno   
\newcount\subsubno   
\newcount\figno
\newcount\tableno

\def\section#1
   {\vskip0pt plus.1\vsize\penalty-250
    \vskip0pt plus-.1\vsize\vskip18pt plus9pt minus6pt
     \subno=0 \exno=0 \figno=0  \eqnno=0
   {\parindent=30pt\raggedright
    \global\advance\secno by 1
    \item{\hbox to 25pt{\large\the\secno\hfill}}\large #1.
    \medskip}}
\def\subsection#1
   {\vskip-\lastskip
    \vskip15pt plus4pt minus4pt
    \bigbreak
    \global\advance\subno by 1
    \noindent {\bf \the\secno.\the\subno\enskip #1. }}

\def\subsubsection#1
   {\vskip-\lastskip
    \exno=0 \caseno=0
     \vskip4pt plus2pt minus2pt
    \bigbreak \global\advance\subsubno by 1
   \noindent {\sl \the\secno.\the\subno.\the\subsubno\enskip #1. }}

\def\boldrule#1{\vbox{\hrule height1pt width#1}}

\def\bline#1{\boldrule{#1truein}}

\def\Table#1#2#3{\vskip-\lastskip
    \vskip4pt plus2pt minus2pt
    \bigbreak \global\advance\tableno by 1
		 \vbox{\centerline{\bf Table \the\secno.\the\subno%
		 \uppercase\expandafter{\romannumeral\the\tableno}}\smallskip
		      \centerline{\sl #1}
 {$$\vbox{\offinterlineskip\tabskip=0pt
       \bline{#2}
       \halign to#2truein{#3}
       \bline{#2}}
   $$}}}

\newbox\strutbox
\setbox\strutbox=\hbox{\vrule height10pt depth 4.5pt width0pt}

\def\defn#1{\vskip-\lastskip
    \vskip4pt plus2pt minus2pt
    \bigbreak
    \global\advance\exno by 1
    \noindent {{\bf Definition\ \the\secno.\the\subno.\the\exno}\enskip
{\rm#1\/}}\smallskip}
\def\example#1{\vskip-\lastskip
    \vskip4pt plus2pt minus2pt \caseno=0
    \bigbreak \global\advance\exno by 1
    \noindent {{\bf Example\ \the\secno.\the\subno.\the\exno}
\enskip{\rm#1}}\smallskip}
\def\exercise#1{\vskip-\lastskip
    \vskip4pt plus2pt minus2pt \caseno=0
    \bigbreak \global\advance\exno by 1
    \noindent {{\bf Exercise\ \the\secno.\the\subno.\the\exno}
\enskip{\rm#1}}\smallskip}
\def\thm#1{\vskip-\lastskip
    \vskip4pt plus2pt minus2pt \caseno=0
    \bigbreak \global\advance\exno by 1
    \noindent {{\bf Theorem\ \the\secno.\the\subno.\the\exno}\enskip
    {\sl#1\/}}\smallskip}
\def\lem#1{\vskip-\lastskip
    \vskip4pt plus2pt minus2pt \caseno=0
    \bigbreak \global\advance\exno by 1
    \noindent {{\bf Lemma\ \the\secno.\the\subno.\the\exno}\enskip
    {\it#1\/}}\smallskip}
\def\remark{\vskip-\lastskip
    \vskip4pt plus2pt minus2pt \caseno=0
    \bigbreak \global\advance\exno by 1
    \noindent {{\bf Remark\ \the\secno.\the\subno.\the\exno}\enskip}}
\def\remarks{\vskip-\lastskip
    \vskip4pt plus2pt minus2pt \caseno=0
    \bigbreak \global\advance\exno by 1
    \noindent {{\bf Remarks\ \the\secno.\the\subno.\the\exno}\enskip}}

\def\=#1{{\bf\bar{\mit#1}}}
\def\^#1{{\bf\hat{\mit#1}}}
\def\~#1{{\bf\tilde{\mit#1}}}
\def\star#1{\item{$\bullet$\ }\underbar{\sl #1}}
\def\pd#1#2{{\partial#1\over\partial#2}}

\def\cc#1{\kappa_{#1}}
\def\pr#1{\mathop{\rm pr}\nolimits^{(#1)}}
\def\hbb#1#2#3{\qquad{\hbox to 100pt{$#1$\hfill}}{\hbox to
200pt{$#2$\hfill}}\hfill#3}
\def\ra{{\hbox to 100pt{\hfill}}\Rightarrow\qquad}

\newcount\remno
\def\tfr#1#2{{\textstyle{#1\over#2}}}
\def\rem#1{\global\advance\remno by 1
\item{\the\remno.\enskip}#1}

\newcount\eqnno
\def\sen{\the\secno}
\def\eqn#1{\global\advance\eqnno by 1
	   \eqno(\sen.\the\eqnno)
	   \expandafter \xdef\csname #1\endcsname
	   {\sen.\the\eqnno}\relax }
\def\eqnn#1{\global\advance\eqnno by 1
	   (\sen.\the\eqnno)
	   \expandafter \xdef\csname #1\endcsname
	   {\sen.\the\eqnno}\relax }
\def\eqnm#1#2{\global\advance\eqnno by 1
	   (\sen.\the\eqnno\hbox{#2})
	   \expandafter \xdef\csname #1\endcsname
	   {\sen.\the\eqnno}\relax }
\def\eqnr#1{(\sen.\the\eqnno\hbox{#1})}

\def\cite#1{[#1]}
\def\DELTA{\Delta}
\def\maple{{\sc maple}}
\def\dgb{{DGB}}
\def\star#1{\item{$\bullet$\ }\underbar{\sl #1}}
\def\bdot#1{{\bf\dot{\mit#1}}}

\def\ft{\bdot{f}}
\def\dgbs{{DGBs}}

\def\pd#1{\partial_{#1}}
\def\fft#1{{\d f_{#1}\over \d t}}
\def\fftt#1{{\d^2 f_{#1}\over \d t^2}}

\font\nrm=cmr9
\font\nbf=cmbx9
\font\nit=cmti9
\font\nsl=cmsl9

\font\ntt=cmtt9

\def\refpp#1#2#3{}{}{}
\def\refjl#1#2#3#4#5{}{}{}{}{}
\def\refbk#1#2#3#4{}{}{}{}
\def\refcf#1#2#3#4#5#6{}{}{}{}{}{}
\newcount\refno
\refno=0
\def\refn#1{\global\advance\refno by 1
\expandafter \xdef\csname #1\endcsname {\the\refno}\relax}

\def\hide#1{}
\def\akns{SWW-AKNS}
\def\hs{SWW-HS}

\refn{refAC}
\refbk{M.J. Ablowitz and P.A. Clarkson}{1991}{Solitons, Nonlinear
Evolution
Equations and Inverse Scattering}{{\fit L.M.S. Lect. Notes
Math.\/}, {\nbf 149}, C.U.P., Cambridge}
\refn{refAKNS}
\refjl{M.J. Ablowitz, D.J. Kaup, A.C. Newell and H. Segur}{1974}{Stud.
Appl.
Math.}{53}{249--315}
\refn{refARSa}
\refjl{M.J. Ablowitz, A. Ramani and H. Segur}{1978}{\PRL}{23}{333--338}
\refn{refARSb}
\refjl{M.J. Ablowitz, A. Ramani and H. Segur}{1980}{\jmp}{21}{715--721}
\refn{refASH}
\refjl{M.J. Ablowitz, C. Schober and B.M.
Herbst}{1993}{\PRL}{71}{2683--2686}
\refn{refVA}
\refjl{M.J. Ablowitz and J. Villarroel}{1991}{\sam}{85}{195--213}
\refn{refAndIb}
\refbk{R.L. Anderson and N.H. Ibragimov}{1979}{Lie-B\"acklund
Transformations in Applications}{SIAM, Philadelphia}
\refn{refBBM}
\refjl{T.B. Benjamin, J.L. Bona and J. Mahoney}{1972}{Phil. Trans. R.
Soc.
Lond. Ser. A}{272}{47--78}
\refn{refBCa}
\refjl{G.W. Bluman and J.D. Cole}{1969}{J. Math. Mech.}{18}{1025--1042}
\refn{refBK}
\refbk{G.W. Bluman and S. Kumei}{1989}{Symmetries and Differential
Equations}{{\fit Appl. Math. Sci.\/} {\nbf 81}, Springer-Verlag,
Berlin}
\refn{refBogi}
\refjl{O.I. Bogoyaviemskii}{1990}{Math. USSR Izves.}{34}{245--259}
\refn{refBogii}
\refjl{O.I. Bogoyaviemskii}{1990}{Russ. Math. Surv.}{45}{1--86}
\refn{refBLMP}
\refjl{M. Boiti, J.J-P. Leon, M. Manna and F.
Pempinelli}{1986}{\IP}{2}{271--279}
\refn{refBuchi}
\refcf{B. Buchberger}{1988}{Mathematical Aspects of Scientific
Software}{Ed.\
J.\ Rice}{Springer Verlag}{pp59--87}
\refn{refCHW}
\refjl{B. Champagne, W. Hereman and P. Winternitz}{1991}{Comp. Phys.
Comm.}{66}{319--340}
\refn{refPACrev}
\refpp{P.A. Clarkson}{1994}{``Nonclassical symmetry reductions for the
Boussinesq equation'', {\fit Chaos, Solitons \&\ Fractals\/}, to
appear}
\refn{refCK}
\refjl{P.A. Clarkson and M.D. Kruskal}{1989}{\jmp}{30}{2201--2213}
\refn{refCMa}
\refjl{P.A. Clarkson and E.L. Mansfield}{1994}{Physica D}{70}{250--288}
\refn{refCMc}
\refjl{P.A. Clarkson and E.L.
Mansfield}{1994}{Nonlinearity}{7}{975--1000}
\refn{refCMb}
\refpp{P.A. Clarkson and E.L. Mansfield}{1994}{``Algorithms for the
nonclassical method of symmetry reductions'', {\fit SIAM J. Appl.
Math.\/}, to
appear}
\refn{refCMd}
\refpp{P.A. Clarkson and E.L. Mansfield}{1994}{``Exact  solutions for
some
2+1-dimensional shallow water wave equations'', preprint, Department of
Mathematics, University of Exeter}
\refn{refCole}
\refjl{J.D. Cole}{1951}{Quart. Appl. Math.}{9}{225--236}
\refn{refCM}
\refjl{R. Conte and M. Musette}{1991}{\jmp}{32}{1450--1457}
\refn{refDTT}
\refjl{P. Deift, C. Tomei and E. Trubowitz}{1982}{Commun. Pure Appl.
Math.}{35}{567--628}
\refn{refDGRW}
\refjl{B. Dorizzi, B. Grammaticos, A. Ramani and P.
Winternitz}{1986}{\JMP}{27}{2848--2852}
\refn{refEF}
\refjl{A. Espinosa and J. Fujioka}{1994}{\JPSJ}{63}{1289--1294}
\refn{refGGKM}
\refjl{C.S. Gardner, J.M. Greene, M.D. Kruskal and R. Miura
M}{1967}{Phys. Rev. Lett}{19}{1095--1097}
\refn{refGNW}
\refjl{C.R. Gilson, J.J.C. Nimmo and R.
Willox}{1993}{\pl}{180A}{337--345}
\refn{refFush}
\refjl{W.I. Fushchich}{1991}{Ukrain. Math. J.}{43}{1456--1470}
\refn{refHere}
\refjl{W. Hereman}{1994}{Euromath Bull.}{1 {\nrm no. 2}}{45--79}
\refn{refHiet}
\refcf{J. Hietarinta}{1990}{Partially Integrable Evolution Equations in
Physics}{Eds. R. Conte and N. Boccara}{{\nit NATO ASI Series C:
Mathematical and Physical Sciences\/}, {\nbf 310}, Kluwer,
Dordrecht}{pp459--478}
\refn{refHirt}
\refcf{R. Hirota}{1980}{Solitons}{Eds. R.K. Bullough and P.J.
Caudrey}{{\nit Topics in Current Physics\/}, {\nbf 17},
Springer-Verlag, Berlin}{pp157--176}
\refn{refHI}
\refjl{R. Hirota and M. Ito}{1983}{\JPSJ}{52}{744--748}
\refn{refHS}
\refjl{R. Hirota and J. Satsuma}{1976}{J. Phys. Soc.
Japan}{40}{611--612}
\refn{refHopf}
\refjl{E. Hopf}{1950}{Commun. Pure Appl. Math.}{3}{201--250}
\refn{refInce}
\refbk{E.L. Ince}{1956}{Ordinary Differential Equations}{Dover, New
York}
\refn{refJM}
\refjl{M. Jimbo and T. Miwa}{1983}{Publ. R.I.M.S.}{19}{943--1001}
\refn{refLU}
\refjl{S.B. Leble and N.V. Ustinov}{1994}{\IP}{210}{617--633}
\refn{refLW}
\refjl{D. Levi and P. Winternitz}{1989}{\jpa}{22}{2915--2924}
\refn{refMD}
\refpp{E.L. Mansfield}{1993}{``{\ntt diffgrob2}: A symbolic algebra
package
for analysing systems of PDE using Maple", {\ntt ftp
euclid.exeter.ac.uk},
login: anonymous, password: your email address, directory: {\ntt
pub/liz}}
\refn{refMF}
\refpp{E.L. Mansfield and E.D.\ Fackerell}{1992}{``Differential
Gr\"obner
Bases",  preprint {\nbf 92/108}, Macquarie University, Sydney,
Australia}
\refn{refMcLO}
\refjl{J.B. McLeod and P.J. Olver}{1983}{SIAM J. Math.
Anal.}{14}{488--506}
\refn{refMLD}
\refjl{M. Musette, F. Lambert and J.C.
Decuyper}{1987}{\JPA}{20}{6223--6235}
\refn{refOlver}
\refbk{P.J. Olver}{1993}{Applications of Lie Groups to Differential
Equations}{Second Edition, {\fit Graduate Texts  Math.} {\nbf
107}, Springer-Verlag, New York}
\refn{refORi}
\refjl{P.J. Olver and P. Rosenau}{1986}{\pl}{114A}{107--112}
\refn{refORii}
\refjl{P.J. Olver and P. Rosenau}{1987}{SIAM J. Appl.
Math.}{47}{263--275}
\refn{refPer}
\refjl{H. Peregrine}{1966}{J. Fluid Mech.}{25}{321--330}
\refn{refReida}
\refjl{G.J. Reid}{1990}{\jpa}{23}{L853--L859}
\refn{refReidb}
\refjl{G.J. Reid}{1991}{Europ. J. Appl. Math.}{2}{293--318}
\refn{refRW}
\refpp{G.J. Reid and A. Wittkopf}{1993}{``A Differential Algebra
Package
for Maple'', {\ntt ftp 137.82.36.21} login: anonymous, password: your
email address, directory: {\ntt pub/standardform}}
\refn{refSchw}
\refjl{F. Schwarz}{1992}{Computing}{49}{95--115}
\refn{refTP}
\refjl{K.M. Tamizhmani and P. Punithavathi}{1990}{\JPSJ}{59}{843--847}
\refn{refTop}
\refjl{V.L. Topunov}{1989}{Acta Appl. Math.}{16}{191--206}
\refn{refWeiss}
\refjl{J. Weiss}{1983}{\jmp}{24}{1405--1413}
\refn{refWTC}
\refjl{J. Weiss, M. Tabor and G. Carnevale}{1983}{\jmp}{24}{522--526}
\refn{refWW}
\refbk{E.E. Whittaker and G.M. Watson}{1927}{Modern Analysis}{Fourth
Edition,
C.U.P., Cambridge}
\refn{refWint}
\refcf{P. Winternitz}{1993}{Integrable Systems, Quantum Groups, and
Quantum Field
Theories}{Eds. L.A. Ibort and M.A. Rodriguez}{{\nit NATO ASI Series C},
{\nbf 409}, Kluwer, Dordrecht}{pp425--495}

\title{Symmetry Reductions and Exact Solutions of Shallow Water Wave
Equations}
\author{Peter A.\ Clarkson and Elizabeth L.\ Mansfield}
\address{Department of Mathematics, University of Exeter, Exeter, EX4
4QE, U.K.}
\bigskip
\date

\beginabstract
In this paper we study symmetry reductions and exact solutions of the
shallow
water wave (SWW) equation
$$u_{xxxt} + \alpha u_x u_{xt} + \beta u_t u_{xx}
- u_{xt} - u_{xx} = 0,\eqno(1)$$
where $\alpha$ and $\beta$ are arbitrary, nonzero, constants, which is
derivable
using the so-called Boussinesq approximation. Two special cases of this
equation, or the equivalent nonlocal equation obtained by setting
$u_x=U$, have
been discussed in the literature. The case $\alpha=2\beta$ was
discussed by
Ablowitz, Kaup, Newell and Segur [{\it Stud.\ Appl.\ Math.}, {\bf53}
(1974)
249], who showed that this case was solvable by inverse scattering
through a
second order linear problem. This case and the case $\alpha=\beta$ were
studied
by Hirota and Satsuma [{\it J.\ Phys.\ Soc.\ Japan}, {\bf40} (1976)
611] using
Hirota's bi-linear technique. Further the case $\alpha=\beta$ is
solvable by
inverse scattering through a third order linear problem.

In this paper a catalogue of symmetry reductions is obtained using the
classical Lie method and the nonclassical method due to Bluman and Cole
[{\it
J.\ Math.\ Mech.\/}, {\bf 18} (1969) 1025]. The classical Lie method
yields
symmetry reductions of (1) expressible in terms of the first, third
and fifth \p\ transcendents and Weierstrass elliptic functions. The
nonclassical
method yields a plethora of exact solutions of (1) with $\alpha=\beta$
which
possess a rich variety of qualitative behaviours. These solutions all
like a
two-soliton solution for $t<0$ but differ radically for $t>0$ and may
be viewed
as a nonlinear superposition of two solitons, one travelling to the
left with
arbitrary speed and the other to the right with equal and opposite
speed.
These families of solutions have important implications with regard to
the
numerical analysis of SWW and suggests that solving (1) numerically
could pose
some fundamental difficulties. In particular, one would not be able to
distinguish the solutions in an initial value problem since an
exponentially
small change in the initial conditions can result in completely
different
qualitative behaviours.

We compare the two-soliton solutions obtained using the nonclassical
method to
those obtained using the singular manifold method and Hirota's
bi-linear method.

Further, we show that there is an analogous nonlinear superposition of
solutions for two $2+1$-dimensional generalisations of the SWW equation
(1)
with $\alpha=\beta$. This yields solutions expressible as the sum of
two
solutions of the Korteweg-de Vries equation.

\endabstract

\vfill\eject

\pageno=1
\headline={\ifodd\pageno\rightheadline\else\leftheadline\fi}
\def\rightheadline{\tenrm\hfil {\sit Nonclassical symmetries and exact
solutions of a shallow water wave equation}\hfil\folio}
\def\leftheadline{\tenrm\hfil {\sit Peter A.\ Clarkson and Elizabeth
L.\ Mansfield}\hfil\folio}

\section{Introduction}
In this paper we discuss symmetry reductions and exact solutions for
the
shallow water wave ({SWW}) equation
$$\DELTA \equiv u_{xxxt} + \alpha u_x u_{xt} + \beta u_t u_{xx}
- u_{xt} - u_{xx} = 0,\eqn{eqgsww}$$
where $\alpha$ and $\beta$ are arbitrary, nonzero, constants, which can
be
derived from the classical shallow water theory in the Boussinesq
approximation
(cf., \cite{\refEF}). This equation may be written in the nonlocal form
(set
$u_x=U$)
$$U_{xxt} + \alpha U U_{t} - \beta U_{x}\partial_x^{-1}U_t - U_{t} -
U_{x} =
0,\eqn{eqgswwv}$$
where $\left(\partial_x^{-1} f\right)(x) = \int_x^\infty f(y)\,\d y$;
in fact
this is the form of the physical model derived in \cite{\refEF}.

Two special cases of (\eqgswwv) have attracted some attention in
the literature, namely $\alpha=2\beta$ and $\alpha=\beta$.
In their seminal paper on soliton theory, Ablowitz, Kaup, Newell and
Segur
\cite{\refAKNS} show that
$$U_{xxt} + 2\beta U U_{t} - \beta U_{x}\partial_x^{-1}U_t -
U_{t} - U_{x} = 0,\eqn{eqswwb}$$
which is (\eqgswwv) with $\alpha=2\beta$ is solvable by inverse
scattering.
Further, Ablowitz \etal \cite{\refAKNS} remark that (\eqswwb) reduces
in the
long wave, small amplitude limit to the celebrated Korteweg-de Vries
(KdV)
equation
$$u_{t} + u_{xxx} + 6uu_x=0,\eqn{eqkdv}$$
which was the first equation to be solved by inverse scattering
\cite{\refGGKM},
and they also comment that (\eqswwb) also has the desirable properties
of the
regularized long wave (RLW) equation \cite{\refBBM,\refPer}
$$u_{xxt} + u u_{x} - u_{t} - u_{x} = 0,\eqn{eqrlw}$$ sometimes known
as the
Benjamin-Bona-Mahoney equation, in that it responds feebly to short
waves.
Additionally, we note that (\eqrlw) and (\eqswwb) have the same linear
dispersion relation $\omega(k)=-k/(1+k^2)$ for the complex exponential
$u(x,t)\sim\exp\{\i[kx+\omega(k)t]\}$. However, in contrast to
(\eqswwb), the
RLW equation (\eqrlw) is thought \underbar{\it not} to be solvable by
inverse
scattering (cf., \cite{\refMcLO}). Hirota and Satsuma \cite{\refHS}
studied both
(\eqswwb) and
$$U_{xxt} + \beta U U_{t} - \beta U_{x}\partial_x^{-1}U_t - U_{t} -
U_{x} =
0,\eqn{eqswwa}$$
i.e.\ (\eqgswwv) with $\alpha=2\beta$, using Hirota's bi-linear method
\cite{\refHirt} and obtained $N$-soliton solutions for both equations
(see
also \cite{\refMLD}).

In the sequel, we shall refer to (\eqgsww) with $\alpha=2\beta$
$$u_{xxxt} + 2\beta u_x u_{xt} + \beta u_t u_{xx} - u_{xt} - u_{xx} =
0,
\eqn{eqswwii}$$
as the {\akns} equation and (\eqgsww) with
$\alpha=\beta$
$$u_{xxxt} + \beta u_x u_{xt} + \beta u_t u_{xx} - u_{xt} - u_{xx} = 0,
\eqn{eqswwi}$$
as the {\hs} equation.

The scattering problem for the {\akns} equation (\eqswwii) is the
second order
problem
\cite{\refAKNS} $$\psi_{xx} + \tfr12\beta u_x \psi =
\lambda\psi,\eqn{scpiia}$$
with associated time-dependence
$$(4\lambda-1)\psi_t = (1- \beta u_t)\psi_x +\tfr12\beta
u_{xt}\psi,\eqn{scpiib}$$ where $\lambda$
is the constant eigenvalue, and $\psi_{xxt}=\psi_{txx}$ if and only if
$u$ satisfies (\eqswwii) We note that (\scpiia) is the time-independent
Schr\"odinger equation which is also the scattering
problem for the KdV equation (\eqkdv) \cite{\refGGKM}. In
contrast, the scattering problem for
the {\hs} equation (\eqswwi) is the third order problem
\cite{\refCM}
$$\psi_{xxx} + \left(\tfr12\beta u_x - 1\right)\psi_x =
\lambda\psi,\eqn{scpia}$$
with associated time-dependence
$$3\lambda\psi_t = (1- \beta u_t)\psi_{xx} + \beta
u_{xt}\psi_x.\eqn{scpib}$$
We remark that (\scpia) is similar to the scattering problem
$$\psi_{xxx} + \tfr14(1+6u)\psi_x +\tfr34\left[u_x -
\i\sqrt{3}\,\partial_x^{-1}(v_t)\right]\psi=
\lambda\psi\eqn{bqspa}$$ which is the scattering problem for the
Boussinesq equation
$$u_{xxxx} + 3(u^2)_{xx} + u_{xx} = u_{tt},\eqn{eqbq}$$
and which has been comprehensively studied by Deift, Tomei and
Trubowitz
\cite{\refDTT}.

The {SWW} equation (\eqgsww) was also discussed by Hietarinta
\cite{\refHiet}
who shows that it can be expressed in Hirota's bi-linear form
\cite{\refHirt} if
and only if either (i), $\alpha=\beta$, when it reduces to (\eqswwi),
or (ii),
$\alpha=2\beta$, when it reduces to (\eqswwii). Further, in
\cite{\refCMc} it
is shown that the {SWW} equation (\eqgsww) satisfies the necessary
conditions
of the \p\ tests due to Ablowitz, Ramani and Segur
\cite{\refARSa,\refARSb} and
Weiss, Tabor and Carnevale \cite{\refWTC} to be completely integrable
if and
only if either $\alpha=\beta$ or $\alpha=2\beta$. These results
strongly suggest
that the {SWW} equation (\eqgsww) is completely integrable if and only
if it
has one of the two special forms ({\eqswwii}) or ({\eqswwi}), which are
both
known to be solvable by inverse scattering.

The classical method for finding symmetry reductions of
partial differential equa\-tions (\pdes) is the Lie group method of
infin\-ites\-imal transformations (cf., \cite{\refBK,\refOlver}).
Though this
method is entirely algorithmic, it often involves a large amount of
tedious
algebra and auxiliary calculations which can become virtually
unmanageable if
attempted manually, and so symbolic manipulation programs have been
developed,
for example in {\sc macsyma}, {\maple}, {\sc mathematica}, {\sc mumath}
and {\sc
reduce}, to facilitate the calculations. An excellent survey of the
different
packages presently available and a discussion of their strengths and
applications is given by Hereman \cite{\refHere}.

In recent years the nonclassical method due Bluman and Cole
\cite{\refBCa} (in
the sequel referred to as the {\it nonclassical method\/}) and the
direct method
of Clarkson and Kruskal \cite{\refCK} have been used to generate many
new
symmetry reductions and exact solutions for several physically
significant
\pdes\ that are {\it not\/} obtainable using the classical Lie method,
which
represents important progress (cf., \cite{\refPACrev,\refFush} and
references therein).

Symmetry groups and associated reductions and exact solutions have
several
different important applications in the context of differential
equations (see,
for example, \cite{\refBK,\refPACrev,\refOlver} for further details and
references): {\par
\star{Derive new solutions from old solutions}. Applying the symmetry
group of
a differential equation to a known solution yields a family of new
solutions.
Quite often interesting solutions can be obtained from trivial ones.
\star{Integration of \odes}. Symmetry groups of \odes\ can be used to
reduce
the order of the equation, for example to reduce a second order
equation
to first order.
\star{Reductions of \pdes}. Symmetry groups of \pdes\ are used to
reduce the
total number of dependent and independent variables, for example from a
\pde\
with two independent and one dependent variables to an \ode.
\star{Linearisation of \pdes}. Symmetries groups can be used to
discover whether
or not a \pde\ can be linearised and to construct an explicit
linearisation when
one exists.
\star{Classification of equations}. Symmetry groups can be used to
classify
differential equations into equivalence classes and to choose simple
representatives of such classes.
\star{Asymptotics of solutions of \pdes}. It is known that as solutions
of
\pdes\ asymptotically tend to solutions of lower-dimensional equations
obtained
by symmetry reduction, some of these special solutions will illustrate
important
physical phenomena. In particular, exact solutions arising from
symmetry methods
can often be effectively used to study properties such as asymptotics
and
``blow-up''.
\star{Numerical methods and testing computer coding}. Symmetry groups
and exact
solutions of physically relevant \pdes\ are used in the design, testing
and
evaluation of numerical algorithms; these solutions provide an
important
{practical} check on the accuracy and reliability of such integrators.
\star{Conservation Laws}. The application of symmetries to conservation
laws
dates back to the work of Noether who proved the remarkable
result that for systems arising from a variational principle, every
conservation
law of the system comes from a corresponding symmetry property.
\star{Further Applications}. There are several other important
applications of
symmetry groups including bifurcation theory, control theory, special
function
theory, boundary value problems and free boundary problems.
\par}

The method used to find solutions of the determining equations for the
infinitesimals in both the classical and nonclassical case is that of
Differential Gr\"obner Bases (\dgbs), defined to be a basis ${\cal B}$
of the
differential ideal generated by the system such that every member of
the ideal
pseudo-reduces to zero with respect to ${\cal B}$. This method provides
a
systematic framework for finding integrability and compatibility
conditions of
an overdetermined system of \pdes. It avoids the
problems of infinite loops in reduction processes, and yields, as far
as is
currently possible, a ``triangulation" of the system from which the
solution set
can be derived more easily (cf.,
\cite{\refCMa,\refMF,\refReida,\refReidb}). In
a sense, a \dgb\ provides the maximum amount of information possible
using
elementary differential and algebraic processes in a finite time.

The triangulations of the systems of determining equations for
infinitesimals
arising in the classical and nonclassical methods in this article were
all
performed using the \maple\ package {\tt diffgrob2} \cite{\refMD}. This
package
was written specifically to handle fully nonlinear equations of
polynomial
type; packages such as those in \cite{\refRW,\refSchw,\refTop} have
been
developed for the study of linear equations. All calculations are
strictly
``polynomial", that is, there is no division. Implemented there are the
Kolchin-Ritt algorithm, the differential analogue of Buchberger's
algorithm
\cite{\refBuchi} using pseudo-reduction instead of reduction, and extra
algorithms needed to calculate a \dgb\ (as far as possible using the
current
theory), for those cases where the Kolchin-Ritt algorithm is not
sufficient
\cite{\refMF}. Designed to be used interactively as well as
algorithmically,
the package has proved useful for solving some fully nonlinear systems.
As yet,
however, algorithmic methods for finding the most efficient orderings,
the best
method of choosing the sequence of pairs to be cross-differentiated,
for
deciding when to integrate and read off coefficients of independent
functions in
one of the variables, for finding the best change of coordinates, and
so on, are
still the subject of much investigation.

In \S2 we find the classical Lie group of symmetries and associated
reductions of
(\eqgsww), which are expressible in terms of the first, third and fifth
\p\
transcendents and Weierstrass elliptic functions. Then in \S3 we
discuss the
nonclassical symmetries and reductions of (\eqgsww). In particular, the
nonclassical symmetry reductions obtained for (\eqswwi) generate a wide
variety
of interesting exact analytical solutions of the equations which we
plot using
{\sc maple}. In \S4 we compare the two-soliton solutions generated in
\S3 with
those obtained using the singular manifold method and Hirota's
bi-linear method.
In \S5 we show that there are analogous symmetry reductions and exact
solutions
also occurs for two
$2+1$-dimensional generalisations of the SWW equation (1) with
$\alpha=\beta$.
We discuss our results in \S6.

\section{Classical Symmetry Reductions of the SWW Equation}
To apply the classical method to the {SWW} equation ({\eqgsww}) we
consider the one-parameter Lie group of infinitesimal transformations
in
$(x,t,u)$ given by
$$\eqalignno{
\~{x} &={x}+ \varepsilon {\xi}(x,t,u) + O(\varepsilon^2),
&\eqnm{eqinftr}{a}\cr
\~{t} &= {t} + \varepsilon {\tau}(x,t,u) + O(\varepsilon^2), &\eqnr{b}
\cr
\~{u} &= {u} + \varepsilon {\phi}(x,t,u) + O(\varepsilon^2),
&\eqnr{c}\cr}$$
where $\varepsilon$ is the group parameter. Then one requires that
this transformation leaves invariant the set
$${\cal S}_{\DELTA} \equiv \left\{u(x,t):
\DELTA=0\right\},\eqn{Sdelta}$$ of
solutions of ({\eqgsww}). This yields an overdetermined, linear system
of
equations for the infin\-ites\-imals ${\xi}(x,t,u)$, ${\tau}(x,t,u)$
and
${\phi}(x,t,u)$. The associated Lie algebra of infin\-ites\-imal
symmetries is
the set of vector fields of the form
$$ {\bf v} = \xi(x,t,u){\pd x} + \tau(x,t,u){\pd t} + \phi(x,t,u){\pd
u}.
\eqn{eqvf}$$ Having determined the infinitesimals, the symmetry
variables are
found by solving the characteristic equations
$${{\d}x \over\xi(x,t,u)} = {{\d}t \over\tau(x,t,u)} = {{\d}u
\over\phi(x,t,u)},
\eqn{chareq}$$ which is equivalent to solving the invariant surface
condition
$$\psi\equiv\xi(x,t,u)u_x +\tau(x,t,u)u_t - \phi(x,t,u)=0.\eqn{insc}$$

The set ${\cal S}_{\DELTA}$ is invariant under the transformation
(\eqinftr)
provided that
$$\left.\pr4{\bf
v}\left(\DELTA\right)\right|_{\DELTA=0}=0,\eqn{privth}$$
where $\pr4{\bf v}$ is the fourth prolongation of the vector field
(\eqvf),
which is given explicitly in terms of ${\xi}$, ${\tau}$ and ${\phi}$
(cf.,
\cite{\refOlver}). This yields a system of fourteen determining
equations,
as calculated using the {\sc macsyma} package {\tt symmgrp.max}
\cite{\refCHW}. A triangulation or standard form
\cite{\refCMa,\refReida} of
these determining equations is the following system of eight equations,
$$\eqalignno{&\xi_{u}=0,\qquad \xi_{t}=0,\qquad \xi_{xx}=0,
\qquad\tau_{u}=0,\qquad \tau_{x}=0,\cr &\alpha\phi_{x}-2\xi_{x}=0,
\qquad\beta\phi_{t}-\tau_{t}-\phi_{x}=0,\qquad
\phi_{u}+\xi_{x}=0,\cr}$$
from which we easily obtain the following infinitesimals,
$$ \xi=\cc1 x+\cc2,\qquad \tau=g(t),\qquad
\phi=-\cc1
\left(u-{2x\over\alpha}-{t\over\beta}\right)+{g(t)\over\beta}+\cc3,
\eqn{clinfs}$$ where $g(t)$ is an arbitrary differentiable function and
$\cc1$,
$\cc2$ and $\cc3$ are arbitrary constants. The associated Lie algebra
is spanned
by the vector fields
$${\bf v}_1 = x{\pd x} -
\left(u-{2x\over\alpha}-{t\over\beta}\right){\pd
u},\qquad {\bf v}_2 = {\pd x},\qquad {\bf v}_3 = {\pd u},\qquad {\bf
v}_4(g) =
g(t)\left({\pd t} + \beta^{-1}{\pd u}\right).$$
Solving ({\chareq}) with $\xi$, $\tau$ and $\phi$ given by (\clinfs),
or
equivalently solving ({\insc}), we obtain the following two canonical
symmetry reductions.

\smallskip\noindent{\bf Case 2.1 $\cc1\ne0$}. In this case we set
$g(t) = f(t)/\ft(t)$, $\cc1=1$ and $\cc2=\cc3=0$ (with $\bdot{f}\equiv
\d f/\d
t$). Hence we obtain the symmetry reduction
$$ u(x,t) = f(t) w(z) + {x\over\alpha}+{t\over\beta},\qquad z=xf(t),
\eqn{clsri} $$ where $w(z)$ satisfies
$$ zw''''+4w'''+ (\alpha+\beta)zw'w'' + \beta ww'' +
2\alpha\left(w'\right)^2=0, \eqn{clsriic}$$
where $'=\d/\d z$. It is straightforward to show using the algorithm of
Ablowitz
\etal \cite{\refARSb} that this equation is of \p-type, i.e., its
solutions have
no movable singularities other than poles, only if either (i),
$\alpha=\beta$ or
(ii), $\alpha=2\beta$. These two special cases (\clsriic) are solvable
in terms
of solutions of the third \p\ equation \cite{\refInce}
$${\d^2y\over\d x^2} = {1\over y}\left({\d y\over\d x}\right)^2-
{1\over x}{\d
y\over\d x} + a y^3 + {b y^2+c\over x} + {d\over y},\eqn{eqpiii}$$
and the fifth \p\ equation,
$${\d^2y\over\d x^2} = \left({1\over 2y}+{1\over y-1}\right)\left({\d
y\over\d
x}\right)^2- {1\over x}{\d y\over\d x}+ {(y-1)^2\over x^2}\left(a y +
{b\over
y}\right) + {cy\over x} + {dy(y+1)\over y-1},\eqn{eqpv}$$ with $a$,
$b$, $c$
and $d$ constants (see \cite{\refCMc} for details). Hence the
\p\ Conjecture
\cite{\refARSa,\refARSb} predicts that a necessary condition for
(\eqgsww) to be
completely integrable is that $(\alpha-\beta)(\alpha-2\beta)=0$, i.e.,
only if
(\eqgsww) has one of the two special forms (\eqswwi) or (\eqswwii).

\smallskip\noindent{\bf Case 2.2 $\cc1=0$}. In this case we set $g(t) =
1/\ft(t)$, $\cc2=1$ and $\cc3=-1/\beta$ and obtain the symmetry
reduction
$$ u(x,t) = w(z) + {t/\beta},\qquad z=x-f(t), \eqn{clsrii}
$$ where $W(z)=w'(z)$ satisfies
$$ \left(W'\right)^2 + \tfr13(\alpha+\beta)W^3= AW+B, \eqn{clsriid}$$
with $A$ and $B$ arbitrary constants. This equation is equivalent
to the Weierstrass elliptic function equation
$$ \left(\wp'\right)^2 = 4\wp^3-g_2\wp-g_3, \eqn{clsriie}$$
where $g_2$ and $g_3$ are arbitrary constants \cite{\refWW}.

We remark that the vector field ${\bf v}_4(g)$
shows that ({\eqgsww}) is invariant under the following variable
coefficient
transformation
$$\~x=x,\qquad \~t= g(t),\qquad \~u=u+[g(t)-t]/\beta,\eqn{eqGalTr}$$
i.e., if
$u(x,t)$ is a solution of ({\eqgsww}), then so is $\~u(\~x,\~t)$.
Further, the
associated Lie algebra ${\fg} = \{{\bf v}_4(g)\}$, with
$g(t)\in C^\infty(\Re)$, is a Virasoro algebra since
$$ [{\bf v}_4(g_1),{\bf v}_4(g_2)]={\bf v}_4\left(g_1\bdot{g}_2
-g_2\bdot{g}_1\right),$$
where $[A,B]=AB-BA$ is the standard commutator.
Virasoro algebras frequently arise in the study of symmetry reductions
for
$2+1$-dimensional equations such as the Kadomtsev-Petviashvili and
Davey-Stewartson equations (see, for example,
\cite{\refWint} and the references therein), but not $1+1$-dimensional
equations such as (\eqgsww).

It might be thought that the presence of the arbitrary function $g(t)$
in the
infinitesimals associated with the {SWW} equation ({\eqgsww}) is an
artefact of the fact that we are considering ({\eqgsww}) rather than
(\eqgswwv),
which is the physical equation. In view of the nonlocal term, we write
the
nonlocal equation (\eqgswwv) as the system
$$U_{xxt} + \alpha U U_{t} + \beta V U_{x} - U_{t} - U_{x} = 0,\qquad
V_x =U_t,\eqn{eqgswws}$$
where $\alpha$ and $\beta$ are arbitrary, nonzero constants. To apply
the
classical method to (\eqgswws), we consider the one-parameter Lie group
of infinitesimal transformations in $(x,y,U,V)$ given by
$$\eqalignno{
\~x &= {x}+ \varepsilon {\xi_1}(x,y,U,V) + O(\varepsilon^2), \cr
\~y &= {y}+ \varepsilon {\xi_2}(x,y,U,V) + O(\varepsilon^2), \cr
\~U &= {U} + \varepsilon {\phi_1}(x,y,U,V) + O(\varepsilon^2), \cr
\~V &= {V} + \varepsilon {\phi_2}(x,y,U,V) + O(\varepsilon^2), \cr}$$
where $\varepsilon$ is the group parameter. Solving the resulting
determining
equations yields the infinitesimals
$$\xi = \kappa_1 x + \kappa_2,\qquad \tau = {g(t)},\qquad
\phi_1 = -2\kappa_1(U-{1/\alpha}),\qquad
\phi_2 =-\left(\kappa_1+{\d g\over\d t}\right)(V-{1/\beta}),$$
where $\kappa_1$ and $\kappa_2$ are arbitrary constants and $g(t)$ is
an
arbitrary differentiable function and the associated vector fields are
$${\bf w}_1 =x\pd x -2(U-{1/\alpha})\pd U +
(v-{1/\beta})\pd V,\qquad{\bf w}_2 =\pd x,\qquad
{\bf w}_3(g) = {g(t)}\pd t + {\d g\over\d t}(V-{1/\beta})\pd V.$$
This means that the system (\eqgswws) is invariant under the
transformation
$$\~x=x,\qquad \~t= g(t),\qquad \~U=U,\qquad
\~V={\beta V+\bdot{g}(t)-t\over\beta\bdot{g}(t)}.$$
Hence if $\{U(x,t),V(x,t)\}$ are solutions of ({\eqgswws}), then so are
$\{\~U(\~x,\~t),\~V(\~x,\~t)\}$.

\section{Nonclassical Symmetry Reductions of the SWW Equation}
There have been several generalisations of the classical Lie group
method for
symmetry reductions. Bluman and Cole \cite{\refBCa}, in their study of
symmetry
reductions of the linear heat equation, proposed the so-called
nonclassical
method of group-invariant solutions. This method involves considerably
more
algebra and associated calculations than the classical Lie method. In
fact, it
has been suggested that for some \pdes, the calculation of these
nonclassical
reductions might be too difficult to do explicitly \cite{\refORi},
especially if
attempted manually since the associated determining equations are now
an
overdetermined, {\it nonlinear\/} system. For some equations such as
the KdV
equation (\eqkdv), the nonclassical method does not yield any
additional
symmetry reductions to those obtained using the classical Lie method,
while
there are \pdes\ which do possess symmetry reductions {\it not\/}
obtainable
using the classical Lie group method. It should be emphasised that the
associated vector fields arising from the nonclassical method do not
form a
vector space, still less a Lie algebra, since the invariant surface
condition
(\insc) depends upon the particular reduction. Subsequently, these
methods were
further generalised by Olver and Rosenau \cite{\refORi,\refORii} to
include
``weak symmetries'' and, even more generally, ``side conditions'' or
``differential constraints'', and they concluded that ``the unifying
theme
behind finding special solutions of \pdes\ is not, as is commonly
supposed,
group theory, but rather the more analytic subject of overdetermined
systems
of \pdes''.

In the nonclassical method one requires only the subset of ${\cal
S}_{\DELTA}$
given by
$${\cal S}_{\DELTA,\psi} = \left\{u(x,t):\DELTA(u)
=0,\psi(u)=0\right\},
\eqn{nclset}$$ where ${\cal S}_{\DELTA}$ is as defined in (\Sdelta) and
$\psi=0$ is the invariant surface condition (\insc), is invariant under
the
transformation ({\eqinftr}). The usual method of applying the
nonclassical
method (e.g., as described in \cite{\refLW}), to the {SWW} equation
(\eqgsww) involves applying the prolongation $\pr4{\bf v}$ to the
system of
equations given by (\eqgsww) and the invariant surface condition
(\insc) and
requiring that the resulting expressions vanish for
$u\in{\cal S}_{\DELTA,\psi}$, i.e.,
$$\left.\pr4{\bf
v}\left(\DELTA\right)\right|_{\DELTA=0,\psi=0}=0,\qquad
\left.\pr1{\bf v}(\psi)\right|_{\DELTA=0,\psi=0}=0.\eqn{plpde}$$ It is
easily
shown that
$$\pr1{\bf v}(\psi)=-\left(\xi_uu_x+\tau_uu_t-\phi_u\right)\psi,$$
which
vanishes identically when
$\psi=0$ without imposing any conditions upon
$\xi$, $\tau$ and $\phi$. However as shown in \cite{\refCMb}, this
procedure
for applying the nonclassical method can create difficulties,
particularly when
implemented in symbolic manipulation programs. These difficulties often
arise
for equations such as (\eqgsww) which require the use of differential
consequences of the invariant surface condition (\insc). In
\cite{\refCMb} we
proposed an algorithm for calculating the determining equations
associated with
the nonclassical method which avoids many of the difficulties commonly
encountered; we use that algorithm here.

In the canonical case when $\tau\not\equiv0$ we set $\tau=1$. (We shall
omit the
special case $\tau\equiv0$; in that case one obtains a single condition
for
$\phi$ with 424 summands, and even considering the subcase $\phi_u=0$
leads to an equation more complex than the one we are studying.)
Eliminating $u_t$, $u_{xt}$ and $u_{xxxt}$, in (\eqgsww) using the
invariant
surface condition (\insc) yields
$$\eqalignno{{\bf\tilde{\hbox{$\Delta$}}}\equiv
\phi_{u}&u_{xxx}+3\phi_{uu}u_x
u_{xx}+3\phi_{ux}u_{xx} +\phi_{uuu}u_x^3+3\phi_{xuu}
u_x^2+3\phi_{xxu}u_x +
\phi_{xxx} \cr &-\xi_{xxx}u_x
-3\xi_{xx}u_{xx}-3\xi_{x}u_{xxx}-\xi_{uuu}u_x^4-3\xi_{xuu}u_x^3-
6\xi_{uu}u_{x}^2u_{xx}\cr &-3\xi_{xxu}u_x^2-9\xi_{xu}u_x
u_{xx}-\xi_{u}\left(4u_xu_{xxx}+3u_{xx}^2\right)-\xi u_{xxxx}\cr
&+\left(\alpha
u_x-1\right)\left[\phi_x + \phi_uu_x-\xi_x u_x-\xi_u u_x^2-\xi
u_{xx}\right]
+u_{xx}\left[\beta \left(\phi-\xi u_x\right)-1\right]
=0,&\eqnn{swwisc}\cr}$$
with $t$ a parameter, which involves the infinitesimals $\xi$ and
$\phi$ that
are to be determined. Now we apply the classical Lie algorithm to this
equation
using the fourth prolongation $\pr4{\bf v}$ and eliminate $u_{xxxx}$
using
({\swwisc}). This yields the following overdetermined, nonlinear system
of
equations for ${\xi}$ and ${\phi}$ (contrast the classical case
discussed in \S2
above where they are linear).
$$\eqalignno{
&\xi_{u}=0,&\eqnm{noncldeqs}{i}\cr
&\phi_{uu}=0,&\eqnr{ii}\cr
&(\alpha + \beta) (\phi_{u} + \xi_{x})=0,&\eqnr{iii}\cr
&\xi \phi_{tu} + 3 \xi^{2} \xi_{xx} + 3 \xi_{t} \xi_{x} - 3 \xi
\xi_{xt} - 3
\xi^{2}\phi_{xu} - \xi_{t} \phi_{u}=0,&\eqnr{iv}\cr
& \alpha \xi \phi_{u}^{2} - \alpha \xi_{t} \phi_{u} + \alpha \xi
\phi_{tu}
- 2 \beta \xi^{2}\phi_{xu} - \alpha \xi^{2} \phi_{xu} + \beta \xi^{2}
\xi_{xx}
- \alpha \xi \xi_{x}^{2} + \alpha \xi_{t} \xi_{x} - \alpha \xi
\xi_{xt}=0,
&\eqnr{v}\cr
& 3 \xi \phi_{xu} \phi_{xx} + \beta \xi \phi \phi_{xx} - \xi_{t}
\phi_{xxx}
- 3 \xi \xi_{xx} \phi_{xx} - \xi \phi_{xx} + \alpha \xi \phi_{x}^{2}
+ 3 \xi \phi_{xxu} \phi_{x} - \xi \xi_{xxx} \phi_{x}\cr
&\qquad- 2 \xi \xi_{x} \phi_{x} + \xi_{t} \phi_{x} + \xi \phi
\phi_{xxxu}
 - \xi \phi \phi_{xu} + \xi \phi_{xxxt} - \xi \phi_{xt}=0,&\eqnr{vi}\cr
& 3 \xi \phi_{u} \phi_{xu} - 6 \xi \xi_{x} \phi_{xu} - 3 \xi_{t}
\phi_{xu} -
\alpha \xi^{2} \phi_{x} - 3 \xi^{2} \phi_{xxu} + \beta \xi \phi
\phi_{u}
- 3 \xi \xi_{xx} \phi_{u} + 3 \xi \phi_{xtu} + \beta \xi \phi_{t}\cr
&\qquad+ \beta \xi \xi_{x} \phi - \beta \xi_{t} \phi + \xi^{2}
\xi_{xxx}
+ 6 \xi \xi_{x} \xi_{xx} + 3 \xi_{t} \xi_{xx} + 2 \xi^{2} \xi_{x}
- \xi \xi_{x} - 3 \xi \xi_{xxt} + \xi_{t}=0,&\eqnr{vii}\cr
&9 \xi \xi_{xx} \phi_{xu} + \beta \xi^{2}\phi_{xx} + \alpha \xi_{t}
\phi_{x}
+ 3 \xi \xi_{x} \phi_{xxu} - 2 \alpha \xi \phi_{u} \phi_{x}
- 6 \xi \phi_{xu}^{2} - \alpha \xi \phi_{xt} - \xi \xi_{x} \xi_{xxx}\cr
&\qquad+ 2 \xi \phi_{xu} + \xi \xi_{xxx} \phi_{u} - 3 \xi \phi_{u}
\phi_{xxu}
- 3 \xi \xi_{xx}^{2} - 2 \beta \xi \phi \phi_{xu} - 3 \xi \phi_{xxtu}
+ 2 \xi \xi_{x} \phi_{u} - 2 \xi \xi_{x}^{2} - \xi^{2} \phi_{xu}\cr
&\qquad- \xi_{t} \phi_{u} + \xi \phi_{tu} + \xi_{t} \xi_{x} - \xi
\xi_{xt}
+ \beta \xi \xi_{xx} \phi + \xi^{2} \phi_{xxxu} - \alpha \xi \phi
\phi_{xu}\cr
&\qquad- \xi_{t} \xi_{xxx} - \xi \xi_{xx} + \xi \xi_{xxxt}
+ 3 \xi_{t} \phi_{xxu}=0.&\eqnr{viii} \cr}$$
These equations were calculated using the {\sc macsyma} package {\tt
symmgrp.max} \cite{\refCHW}. We then used the method of {DGBs} as
outlined
in \cite{\refCMa,\refCMb} to solve this system.

\smallskip\noindent{\bf Case 3.1 $\alpha+\beta\ne 0$}.
In this case it is straightforward to obtain from (\noncldeqs{i--v}),
the
condition
$$ \xi_{xx} \xi^{2} (\alpha+\beta) (3 \beta - 2 \alpha)=0.$$
The case $3 \beta - 2 \alpha=0$ leads to no solutions different from
those
obtained using the classical method.

\smallskip\noindent{\sl Subcase 3.1.1 $\xi_x\ne 0$}.
This is the generic case which has the solution
$$\xi = {(\cc1 x + \cc2)f(t)},\qquad \phi =- {\cc1f(t)}\left(u - {2
x\over \alpha} + {\cc2 - t\over\beta}\right) + {1\over\beta},$$
where $f(t)$ is an arbitrary differentiable function and $\cc1\not=0$
and
$\cc2$ are arbitrary constants. These are equivalent to the
infinitesimals
({\clinfs}) obtained using the classical method.

\smallskip\noindent{\sl Subcase 3.1.2 $\xi_x= 0$}.
In this case it is easy to obtain the condition
$$\phi_{xx}\xi^3(\beta-\alpha)(\alpha+\beta)=0.$$
There are two subcases to consider.

\noindent(i) $\alpha\ne\beta$, $\xi_x=0$. In this case the solution is
$$\xi= {f(t)},\qquad \phi= {\cc{3}f(t)}+{1/\beta},$$
where $f(t)$ is an arbitrary differentiable function and $\cc3$ is an
arbitrary
constant, which is equivalent to the infinitesimals ({\clinfs})
obtained using
the classical method in the case when $\cc1=0$.

\noindent(ii) $\alpha= \beta$, $\xi_x=0$.
In this case, we obtain the following \dgb\ for $\xi$, $\phi$
$$\eqalignno{
&\xi_u=0,\qquad\xi_x=0,\qquad\phi_u=0,\cr
&\xi\phi_{xxxx}-(\xi+1)\phi_{xx}+\beta\phi\phi_{xx}+\beta\phi_x^2=0,\cr
&\beta\xi^2\phi_x-\beta\xi\phi_t+\beta\xi_t\phi-\xi_t=0.\cr}$$
Solving these yields
$$\xi=\fft{},\qquad\phi=2V(\zeta)\fft{} +{1\over\beta},\eqn{weqn}$$
where $\zeta=x+f(t)$, $f(t)$ is an arbitrary differentiable function
and
$V(\zeta)$ satisfies
$$V_{\zeta\zeta}+\beta V^2-V = \cc{4}\zeta+\cc5,\eqn{eqpi}$$
with $\cc{4}$ and $\cc{5}$ arbitrary constants.
If $\cc{4}\ne0$, then this equation is equivalent to the first
Painlev\'e
equation \cite{\refInce}
$${\d^2y\over\d x^2} = 6y^2 + x,\eqn{pieq}$$
otherwise it is equivalent to the Weierstrass elliptic
function equation ({\clsriie}).

Thus solving the characteristic equations (\chareq) yields the
nonclassical
reduction
$$ u(x,t) = v(\zeta)+w(z)+t/\beta,\qquad \zeta=x+f(t),\quad z=x-f(t),
\eqn{eqstar}$$
where $f(t)$ is an arbitrary function and $V(\zeta)=v_\zeta$ and
$W(z)=w_z$ satisfy
$$ V_{\zeta\zeta} + \beta V^2 -V = - \lambda \zeta + \mu_1,
\eqno\eqnm{VWeq}{a}$$ and
$$ W_{zz} + \beta W^2 - W = \lambda z+ \mu_2,\eqno\eqnr{b}$$
respectively, where $\mu_1$ and $\mu_2$ are arbitrary constants and
$\lambda$ is (effectively) a ``separation'' constant. If
$\lambda\not=0$ then
these equations are equivalent to the first Painlev\'e equation
({\pieq}),
whilst if $\lambda=0$ then they are equivalent to the Weierstrass
elliptic
function equation ({\clsriie}).

In particular, if $\lambda=0$ (we set $\beta=1$ without loss of
generality) then
equations ({\VWeq}) possess the special  solutions
$$ V(\zeta) = \left\{{6\kappa_1^2}\sech^2\left(\kappa_1\zeta\right)
+\tfr12-2\kappa_1^2\right\}, \qquad W(z) =
\left\{6\kappa_2^2\sech^2\left(\kappa_2
z\right)+\tfr12-2\kappa_2^2\right\},
$$ where $\kappa_1=\tfr12\left(1+4\mu_1\right)^{1/4}$ and
$\kappa_2=\tfr12\left(1+4\mu_2\right)^{1/4}$. Hence we obtain the exact
solution of ({\eqswwi}) given by
$$\eqalignno{ u(x,t) =
\left\{6\kappa_1\right.&\tanh\left\{\kappa_1\left[x+f(t)
\right]\right\}
+6\kappa_2\tanh\left\{\kappa_2\left[x-f(t)\right]\right\}\cr&+
{\left.x(1-2\kappa_1^2-2\kappa_2^2) + 2f(t)(\kappa_2^2-\kappa_1^2) +
t\right\}},&\eqnn{swwsol}\cr}
$$ where $f(t)$ is an arbitrary differentiable function.

If $\mu_1=\mu_2=0$ then $\kappa_1=\kappa_2=\tfr12$ (with $\beta=1$) and
(\swwsol) simplifies to
$$u(x,t) =\left\{{3}\tanh\left\{\tfr12\left[x+f(t)\right]\right\}
+ {3}\tanh\left\{\tfr12\left[x-f(t)\right]\right\}+
t\right\}.\eqn{swwsoli}$$
In Figure 1 we plot $u_x$ with $u$ given by (\swwsoli) for various
choices of
the arbitrary function $f(t)$. This is one of the simplest, nontrivial
family
of solutions of (1.1) with $\alpha=\beta(=1)$, using this reduction. In
Figure
1, $f(t)$ is chosen so that $f(t)\sim t+t_0$, as $t\to-\infty$, where
$t_0$ is a
constant. Consequently all the solutions plotted in Figure 1 have a
similar
asymptotic behaviour as $t\to-\infty$. However the  asymptotic
behaviours as
$t\to\infty$ are radically different.

In the special case when $f(t)=ct$, then choosing
$\kappa_1=\tfr12(1+1/c)^{1/2}$ and
$\kappa_2=\tfr12(1-1/c)^{1/2}$ in (\swwsol) yields the two-soliton
solution for
({\eqswwi}) given by
$$ u(x,t) = {3\over\sqrt{c}}\left\{\sqrt{c+1}\,
\tanh\left[\tfr12\sqrt{(1+1/c)}\,(x+ct)\right] + \sqrt{c-1}\,
\tanh\left[\tfr12\sqrt{(1-1/c)}\,(x-ct)\right]\right\}.\eqn{swwsolii}$$
This solution is of special interest since such two-soliton solutions
are
normally associated with so-called Lie-B\"acklund transformations (cf.,
\cite{\refAndIb}), whereas (\swwsolii) has arisen from a Lie point
symmetry,
albeit nonclassical. Plots of (\swwsolii) and its $x$-derivative are
given in
Figure 2(a) for $c=3$. These should be compared with the plots in
Figure 2(b) of the $x$-derivative of solution (4.6) below for the
\hs\ equation
with $\cc1=2$, $\cc2=1.7$ and $\cc1=\tfr34$, $\cc2=\tfr23$ (so that
$A_{12}=1$
in both cases).

We stress that the ``decoupling'' of the nonclassical reduction
(\eqstar) into
a function of $\zeta=x+f(t)$ and a function of $z=x-f(t)$ occurs for
the SWW
equation (\eqgsww) \underbar{\it only} in this special case when
$\alpha=\beta$.

\smallskip\noindent{\bf Case 3.2 $\alpha+\beta=0$}. This case also
leads to no
solutions different from those obtained using the classical method in
\S2.

To conclude this section we briefly consider the equations
$$\eqalignno{
u_{xxxx}+\alpha u_xu_{xt}+\beta
u_tu_{xx}-u_{xt}-u_{xx}&=0,&\eqnn{eqgswwa}\cr
u_{xxtt}+\alpha u_xu_{xt}+\beta
u_tu_{xx}-u_{xt}-u_{xx}&=0,&\eqnn{eqgswwb}\cr
}$$
where $\alpha$ and $\beta$ are arbitrary, nonzero, constants, which are
variants
of the SWW equation (\eqgsww). Both these equations are thought to be
{\it
nonintegrable\/}, i.e.\ not solvable by inverse scattering, for all
values of
$\alpha$ and $\beta$ since it is straightforward to show that neither
satisfies
the \p\ PDE test due to Weiss \etal \cite{\refWTC}.

Applying the classical Lie method to equations (\eqgswwa) and
(\eqgswwb)
yields the infinitesimals
$$\vbox{\halign{$\displaystyle#$,\qquad & $\displaystyle#$,\qquad &
$\displaystyle#$,\cr
\xi = \cc1x+\cc2 & \tau = (3\cc1+\cc3)t+\cc4 & \phi =
(\cc1+\cc3)u-{\cc3 x\over\alpha}+{2\cc1 t\over\beta}+\cc5\cr
\xi = \cc1x+\cc2 & \tau = -(\cc1+\alpha\cc3)t+\cc4 & \phi =
(\cc1+\alpha\cc3)u-\cc3 x-2(\cc1+\alpha\cc3)t/\beta\cr}}$$
respectively, where $\cc1,\cc2,\ldots,\cc5$ are arbitrary constants. In
contrast
to the SWW equation (\eqgsww), these symmetry groups are finite
dimensional.

If $\alpha=\beta$, then equations (\eqgswwa) and (\eqgswwb) both
possess
nonclassical reductions of the form
$$u(x,t) = v(\zeta) + w(z),\qquad \zeta=x+c t,\quad z=x-c t$$
where for (\eqgswwa), $V(\zeta)=v_\zeta$ and $W(z)=w_z$ satisfy
$$\eqalignno{
&V_{\zeta\zeta} - (1+c)V + \beta c V^2 = - \lambda \zeta +
\lambda_1, &\eqnm{eqgswwc}{i}\cr
&W_{zz} + (c-1)W - \beta c W^2 =\lambda z+ \lambda_2,
&\eqnr{ii}\cr}$$
and for (\eqgswwb), $V(\zeta)=v_\zeta$ and $W(z)=w_z$ satisfy
$$\eqalignno{ &c^2V_{\zeta\zeta} - (1+c)V + \beta c V^2 =
- \lambda \zeta + \lambda_1,&\eqnm{eqgswwd}{i}\cr
&c^2W_{zz} + (c-1)W - \beta c W^2 = \lambda z+
\lambda_2,&\eqnr{ii}\cr}$$ with $\lambda$, $\lambda_1$ and $\lambda_2$
arbitrary
constants. If $\lambda\not=0$, then equations (\eqgswwc,\eqgswwd) are
solvable
in terms of the first Painlev\'e equation ({\pieq}), whilst if
$\lambda=0$ then
they are equivalent to the Weierstrass elliptic function equation
({\clsriie}).

Setting $\lambda=\lambda_1=\lambda_2=0$, and solving equations
(\eqgswwc) and
(\eqgswwd) yields the exact solutions of (\eqgswwa) and (\eqgswwb) with
$\alpha=\beta$ given by
$$\eqalignno{ u(x,t) &= {3\over\beta c}
\left\{\sqrt{1+c}\,\tanh\left[\tfr12\sqrt{1+c}\,(x+ct)\right]
-\sqrt{1-c}\,\tanh\left[\tfr12\sqrt{1-c}\,(x-ct)\right]\right\},
&\eqnn{eqgswwasol}\cr
u(x,t) &= {3\over\beta}\left\{\sqrt{1+c}\,
\tanh\left[{\sqrt{1+c}\over2c}\,(x+ct)\right]-\sqrt{1-c}\,
\tanh\left[{\sqrt{1-c}\over2c}\,(x-ct)\right]\right\},
&\eqnn{eqgswwbsol}\cr}$$ respectively, which are analogues of
(\swwsolii).
Plots of the solutions (\eqgswwasol) and (\eqgswwbsol) and their
derivatives with
respect to $x$ are given in Figures 3 and 4, respectively. The
solutions
(\eqgswwasol) and (\eqgswwbsol) are plotted in Figures 3(i) and 4(i)
and look
like the elastic interaction of a ``kink'' and an ``anti-kink''
solution.
The $x$-derivatives of (\eqgswwasol) and (\eqgswwbsol) are plotted in
Figures
3(ii) and 4(ii) and look like the elastic interaction of two
``soliton''
solutions. These plots are very similar to the plots of (\swwsolii) and
its
$x$-derivative given in Figure 2, however whereas (\eqswwi) is
completely
integrable, both (\eqgswwa) and (\eqgswwb) are seemingly
non-integrable. These
solutions (\eqgswwasol) and (\eqgswwbsol) are of particular interest
since such
solutions are normally associated with integrable equations, whereas
they arise
here for nonintegrable equations. Furthermore, like (\swwsolii), these
``two-soliton'' solutions have also arisen from nonclassical
reductions. As
mentioned above, such solutions are normally associated with
Lie-B\"acklund
transformations (cf., \cite{\refAndIb}).

\section{Soliton Solutions of the \hs\ and \akns\ Equations}
\subsection{The \hs\ Equation}
Exact solutions of the {\hs} equation (\eqswwi) can be obtained using
the
so-called singularity manifold method which uses truncated
\p\ expansions
\cite{\refWeiss,\refWTC}. If we seek a solution of (\eqswwi) in the
form
$$u(x,t) = {6\over\beta}\,{\phi_x(x,t)\over\phi(x,t)},\eqn{eqsmexp}$$
and then
equate coefficients of powers of $\phi$ to zero, we find that
$\phi(x,t)$
satisfies the overdetermined system
$$\eqalignno{&\phi_{xxxt} - \phi_{xx} - \phi_{xt}
=0,&\eqnm{smeqs}{a}\cr
&\phi_t\phi_{xxx} - 3\phi_{xt}\phi_{xx} +3\phi_x\phi_{xxt}-
\phi_{x}(\phi_x +
\phi_{t}) =0.&\eqnr{b}\cr}$$
(A DGB analysis of this system leads to some very
complex expressions. Although it does yield some \odes\ in $x$ for
$\phi$ in the various subcases they appear difficult to solve.) Now
suppose we seek a solution of these equations in the form
$$\phi(x,t) = \alpha_1\exp\left\{\kappa_1x+\mu_1t\right\}+
\alpha_2\exp\left\{\kappa_2x+\mu_2t\right\} + \alpha_0,\eqn{smsol}$$
where
$\alpha_0$, $\alpha_1$, $\alpha_2$, $\kappa_1$, $\kappa_2$, $\mu_1$ and
$\mu_2$
are constants such that $\alpha_0\alpha_1\alpha_2\not=0$. It is
straightforward to show that equations (\smeqs) have a solution of the
form
(\smsol) provided that $\mu_1 = \kappa_1/(\kappa_1^2-1)$,
$\mu_2 = \kappa_2/(\kappa_2^2-1)$ and $\kappa_1$ and $\kappa_2$ satisfy
the
constraint
$$\kappa_1^2-\kappa_1\kappa_2+\kappa_2^2=3.\eqn{smcond}$$
Thus we obtain the following exact solution of the {\hs} equation
(\eqswwi) given by
$$u(x,t) ={6\over\beta}\,
{\displaystyle\alpha_1\kappa_1\exp\left\{\kappa_1x+{\kappa_1t/
(\kappa_1^2-1)}\right\}+
\alpha_2\kappa_2\exp\left\{\kappa_2x+{\kappa_2t/(\kappa_2^2-1)}\right\}
\over \displaystyle\alpha_0+\alpha_1\exp\left\{\kappa_1x+
{\kappa_1t/(\kappa_1^2-1)}\right\}+
\alpha_2\exp\left\{\kappa_2x+{\kappa_2t/(\kappa_2^2-1)}\right\}},
\eqn{swwsmsol}$$
provided $\kappa_1$ and $\kappa_2$ satisfy (\smcond). It should be
noted that
(\swwsmsol) and (\swwsolii) are fundamentally different solutions of
the {\hs}
equation (\eqswwi) as we shall now demonstrate.

The general two-soliton solution of (\eqswwi) is given by
$$u(x,t) = {6\over\beta}\,
{\kappa_1\alpha_1\exp\left(\eta_1\right)+\kappa_2\alpha_2\exp\left(\eta_2\right)
+(\kappa_1+\kappa_2)A_{12}\exp\left(\eta_1+\eta_2\right)
\over 1+\alpha_1\exp\left(\eta_1\right)+\alpha_2\exp\left(\eta_2\right)
+A_{12}\exp\left(\eta_1+\eta_2\right)},\eqno\eqnm{swwisol}{a}$$
where
$$\eqalignno{&\eta_1 = \kappa_1x+{\kappa_1t/(\kappa_1^2-1)},\qquad
\eta_2 = \kappa_2x+{\kappa_2t/(\kappa_2^2-1)},&\eqnr{b}\cr
&A_{12} = \alpha_1\alpha_2
{(\kappa_1-\kappa_2)^2(\kappa_1^2-\kappa_1\kappa_2+\kappa_2^2-3)\over
(\kappa_1+\kappa_2)^2(\kappa_1^2+\kappa_1\kappa_2+\kappa_2^2-3)},
&\eqnr{c}\cr}$$
with $\alpha_1$, $\alpha_2$, $\kappa_1$ and $\kappa_2$ arbitrary
constants
\cite{\refHS} (see also \cite{\refMLD}). A straightforward method of
obtaining
this solution is to substitute (\eqsmexp) into {\hs} equation (\eqswwi)
and then
integrate once. This yields the bi-linear equation,
$$\phi\phi_{xxxt} - \phi_t\phi_{xxx} +
3(\phi_{xx}\phi_{xt}-\phi_x\phi_{xxt}) -
\phi(\phi_{xx}+\phi_{xt})+\phi_x(\phi_x+\phi_t)=0,\eqn{hsphieq}$$
where we have set the function of integration to zero. Now, by seeking
a
solution of this equation, in the form
$$\phi(x,t) =
1+\alpha_1\exp\left(\eta_1\right)+\alpha_2\exp\left(\eta_2\right)
+A_{12}\exp\left(\eta_1+\eta_2\right),$$
with $\eta_1 = \kappa_1x+\mu_1t$ and $\eta_2 = \kappa_2x+\mu_2t$, it is
easy to
show that necessarily $\mu_1 = \kappa_1/(\kappa_1^2-1)$, $\mu_2 =
\kappa_2/(\kappa_2^2-1)$ and $A_{12}$ is given by (\swwisol{c}); this
technique
may be viewed as a variant of Hirota's bi-linear method
\cite{\refHirt}. Four
plots of the $x$-derivative of (\swwisol), i.e.\ solutions of
(\eqswwa), are
given in Figure 5 for (i), $\kappa_1=2$, $\kappa_2=1.7$, (ii),
$\kappa_1=\tfr34$,
$\kappa_2=\tfr23$, (iii), $\kappa_1=1.1$,
$\kappa_2=(11+3\sqrt{93}\,)/20$ (so
that $A_{12}=0$), and (iv), $\kappa_1=0.8$,
$\kappa_2=\tfr15(2+3\sqrt{7}\,)$ (so
that $A_{12}=0$). Plots 5(i) and (ii) illustrate ``standard''
two-soliton
interaction whilst plots 5(iii) and (iv) are in the special case when
$A_{12}=0$.  Plots of the $x$-derivative of (\swwisol) for Figure 2(b)
of the
$x$-derivative of solution (4.6) below for the \hs\ equation,
i.e.\ solutions of
(\eqswwa), for $\cc1=2$, $\cc2=1.7$ and $\cc1=\tfr34$, $\cc2=\tfr23$
(so that
$A_{12}=1$ in both cases).

The solution (\swwsmsol) is the special case of (\swwisol) with
$A_{12}=0$;
Hirota and Ito \cite{\refHI} refer to this as being the ``resonant
state'' where
either a single soliton splits into two solitons, see Figure 5(iii), or
two
solitons fuse together after colliding with each other, see Figure
5(iv). On the
other hand, (\swwsolii) is the special case of (\swwisol) with
$A_{12}=1$ where
two solitons pass through each other with no phase shift as a
consequence of the
interaction. Thus whereas both solutions are asymptotically equivalent
as
$t\to-\infty$, they are qualitatively very different as $t\to\infty$.
This shows
that the nonclassical method and the singular manifold method do not,
in
general, yield the same solution set.

We remark that the \hs\ equation (\eqswwi) also possesses the solution
$$u(x,t) = {6\over\beta}\,{\kappa_1[\alpha_1\exp\left(\eta_1\right)-
\alpha_2\exp\left(-\eta_1\right)]-\kappa_2B_{12}\sin\left(\eta_2+\delta_0\right)
\over\alpha_1\exp\left(\eta_1\right)+\alpha_2\exp\left(-\eta_1\right)
+B_{12}\cos\left(\eta_2+\delta_0\right)},\eqno\eqnm{swwisola}{a}$$
where
$$\eqalignno{
\eta_1 &= \kappa_1x+{(\kappa_1^2+\kappa_2^2-1)t\over
(\kappa_1^2-2\kappa_1+\kappa_2^2+1)
(\kappa_1^2+2\kappa_1+\kappa_2^2+1)},&\eqnr{b}\cr
\eta_2 &= \kappa_2x-{(\kappa_1^2+\kappa_2^2+1)t\over
(\kappa_1^2-2\kappa_1+\kappa_2^2+1)
(\kappa_1^2+2\kappa_1+\kappa_2^2+1)},&\eqnr{c}\cr
B_{12}&=2(\alpha_1\alpha_2)^{1/2}{\kappa_1\over\kappa_2}
\left(3\kappa_1^2-\kappa_2^2-3\over
3\kappa_2^2-\kappa_1^2+3\right)^{1/2},&\eqnr{d}\cr}$$ with
$\alpha_1$, $\alpha_2$, $\kappa_1$, $\kappa_2$ and $\delta_0$ arbitrary
constants. Leble and Ustinov \cite{\refLU} show that solutions of the
form
(\swwisola) exist for several of \pdes\ that are solvable by inverse
scattering through third order linear problems. Two plots of the
$x$-derivative
of (\swwisola), i.e.\ solutions of (\eqswwa), are given in Figure 6 for
(i),
$\kappa_1=1$, $\kappa_2=1.17$ and (ii), $\kappa_1=1$, $\kappa_2=1.24$.

\subsection{The \akns\ Equation}
The general two-soliton solution of the \akns\ equation (\eqswwii) is
given by
$$u(x,t) = {6\over\beta}\,{\kappa_1\alpha_1\exp\left(\eta_1\right)+
\kappa_2\alpha_2\exp\left(\eta_2\right)
+(\kappa_1+\kappa_2)A_{12}\exp\left(\eta_1+\eta_2\right)
\over 1+\alpha_1\exp\left(\eta_1\right)+\alpha_2\exp\left(\eta_2\right)
+A_{12}\exp\left(\eta_1+\eta_2\right)},\eqno\eqnm{swwiisol}{a}$$
where
$$\eta_1 = \kappa_1x+{\kappa_1t/(\kappa_1^2-1)},\qquad
\eta_2 = \kappa_2x+{\kappa_2t/(\kappa_2^2-1)},\qquad
A_{12} =\alpha_1\alpha_2
{\left(\kappa_1-\kappa_2\over\kappa_1+\kappa_2\right)^2},\eqno\eqnr{b}$$
with
with $\alpha_1$, $\alpha_2$, $\kappa_1$ and $\kappa_2$ arbitrary
constants
\cite{\refHS} (see also \cite{\refMLD}). Two plots of the
$x$-derivative of
(\swwiisol), i.e.\ solutions of (\eqswwb), are given in Figure 7 for
(i),
$\kappa_1=2$, $\kappa_2=1.7$ and  (ii), $\kappa_1=\tfr34$,
$\kappa_2=\tfr45$.

To apply the singularity manifold method to the \akns\ equation
(\eqswwii), we
seek a solution in the form
$$u(x,t) = {4\over\beta}\,{\phi_x(x,t)\over\phi(x,t)}.\eqn{eqsmexp}$$
Equating coefficients of powers of $\phi$ to zero, yields the
overdetermined
system
$$\eqalignno{
&\phi_{xxxxt} - \phi_{xxx} - \phi_{xxt} =0,&\eqnm{smeqsb}{a}\cr
&\phi_t\phi_{xxxx} +4\phi_{x}\phi_{xxxt} - 2\phi_{xx}\phi_{xxt} -
3\phi_{x}\phi_{xx} -\phi_{t}\phi_{xx}- 2\phi_{x}\phi_{xt}
=0,&\eqnr{b}\cr
&2\phi_t\phi_{x}\phi_{xxx} -2\phi_{x}\phi_{xx}\phi_{xt} +
2\phi_{x}^2\phi_{xxt}
-\phi_{t}(\phi_{x}^2+\phi_{xx}^2) -\phi_{x}^3 =0.&\eqnr{c}\cr}$$
We note that this is a system of three equations in contrast to that
for the
\hs\ equation (\eqswwi) where $\phi(x,t)$ satisfies a system of two
equations
(\smeqs). It is easily shown that there exist \underbar{\it no}
solutions of
(\smeqsb) in the form (\smsol) if $\alpha_0\alpha_1\alpha_2\not=0$.
Additionally substituting (\eqsmexp) into (\eqswwii) yields a
\underbar{\it
tri-linear} equation for $\phi(x,t)$, whereas the analogous equation
for
(\eqswwi), i.e.\ (\hsphieq), is bi-linear.

\section{Nonlinear Superposition for 2+1-dimensional Equations}
In this section we show that nonlinear superposition of solutions can
also
occur for two $2+1$-dimensional equations of the \hs\ equation
(\eqswwi). Two
$2+1$-dimensional generalisations of the SWW equation (\eqgsww) are the
following equations
$$\eqalignno{
u_{yt}+\alpha u_xu_{xy}+\beta u_yu_{xx}-u_{xxxy}&=0,&\eqnn{eqswwIIi}\cr
u_{xt}+\alpha u_xu_{xy}+\beta
u_yu_{xx}-u_{xxxy}&=0.&\eqnn{eqswwIIii}\cr}$$
We remark that these two equations differ only in the first term, which
belongs to the linear part, yet their symmetries differ remarkably.
Further
both (\eqswwIIi) and (\eqswwIIii) reduce to the KdV equation
(\eqkdv) if $y=x$.

Boiti, Leon, Manna and Pempinelli \cite{\refBLMP} developed an inverse
scattering scheme to solve the Cauchy problem for (\eqswwIIi) with
$\alpha=\beta$, i.e.
$$u_{yt} + u_{xxxy} +\beta u_{xx}u_y +\beta u_xu_{xy}=0,\eqn{eqblmp}$$
which is a $2+1$-dimensional generalisation of the \hs\ equation
(\eqswwi),
for initial data decaying sufficiently rapidly at infinity. This
inverse
scattering scheme is formulated as a nonlocal Riemann-Hilbert problem
and
involves a so-called ``weak-Lax pair''. We note that both (\eqblmp) and
\hs\
(\eqswwi) are reductions of the
$3+1$-dimensional equation
$$u_{yt} + u_{xxxy} +\beta u_{xx}u_y +\beta u_xu_{xy}
-u_{xz}=0,\eqn{eqjm}$$
which was introduced by Jimbo and Miwa \cite{\refJM} as the second
equation in
the so-called Kadomtsev-Petviashvili hierarchy, though (\eqjm) is
not completely integrable in the usual sense (cf., \cite{\refDGRW}).
Bogoyaviemskii \cite{\refBogi,\refBogii} discusses the inverse
scattering method of solution for (\eqswwIIii) with $\alpha=2\beta$,
i.e.,
$$u_{xt} + u_{xxxy} +\beta u_{xx}u_y +2\beta
u_xu_{xy}=0,\eqn{eqbogii}$$
which is a $2+1$-dimensional generalisation of the \akns\ equation
(\eqswwii);
it should be noted that the $2+1$-dimensional generalisation of
(\eqswwii)
discussed by Gilson \etal \cite{\refGNW} is different to (\eqswwIIii).

It is routine to show that (\eqswwIIi) satisfies the necessary
conditions
of the Painlev\'e PDE test due to Weiss \etal \cite{\refWTC} to be
completely
integrable if and only if $\alpha=\beta$, i.e.\ when it reduces to
(\eqblmp),
and (\eqswwIIii) satisfies these necessary conditions if and only if
$\alpha=2\beta$, i.e.\ when it reduces to (\eqbogii). This strongly
suggests
that (\eqswwIIi) and (\eqswwIIii) are solvable by inverse scattering
only in
these two special cases, both of which are known to be completely
integrable
(\cite{\refBLMP} and \cite{\refBogi,\refBogii}, respectively).

\subsection{Classical Symmetries}
To apply the classical method to (\eqswwIIi), we consider
the one-parameter Lie group of infinitesimal transformations in
$(x,y,t,u)$
given by
$$\eqalignno{
\~x &= {x}+ \varepsilon {\xi_1}(x,y,t,u) + O(\varepsilon^2), \cr
\~y &= {y}+ \varepsilon {\xi_2}(x,y,t,u) + O(\varepsilon^2), \cr
\~t &= {t} + \varepsilon {\xi_3}(x,y,t,u) + O(\varepsilon^2), \cr
\~u &= {u} + \varepsilon {\phi}(x,y,t,u) + O(\varepsilon^2), \cr}$$
where $\varepsilon$ is the group parameter. Solving the resulting
determining
equations yields the infinitesimals
$$\eqalignno{&\xi_1=\cc1x+f_1(t),\qquad \xi_2=g(y),\qquad \xi_3=3\cc1
t+\cc2,\qquad \phi= -\cc1u +{x\over\alpha}\fft1+f_2(t),\cr}$$
if $\alpha \not=\beta$ and
$$\eqalignno{
&\xi_1=x\fft{1}+f_2(t),\qquad \xi_2=g(y),\qquad \xi_3=3f_1(t),\qquad
\phi=-u\fft1 +{x^2\over2\alpha}\fftt1 +{x\over\alpha}\fft2+f_3(t),
\cr}$$
if $\alpha =\beta$, where $\cc1$, $\cc2$ and $\cc3$ are arbitrary
constants
and $f_1(t)$, $f_2(t)$, $f_3(t)$ and $g(y)$ are arbitrary
differentiable
functions \cite{\refCMd} (classical symmetries in the case $\alpha
=\beta$
are also discussed in \cite{\refTP}).

Applying the classical Lie method to (\eqswwIIii) yields the
infinitesimals
$$\eqalignno{&\xi_1=\cc1x+f_1(t),\qquad
\xi_2=\cc2t+\cc3y+\cc4,\qquad
\xi_3=(2\cc1+\cc3)t+\cc5,\cr
&\phi=-\cc1 u +{y\over\beta}\fft1+{\cc2x\over\alpha} +f_2(t),\cr}$$
if $\alpha \not=2\beta$ and
$$\eqalignno{&\xi_1=(\cc1t+\cc2)x+f_1(t),\qquad \xi_2=(2\cc1 y+\cc3)t
+\cc4y+\cc5,\qquad
\xi_3=2\cc1 t^2+(2\cc2+\cc4)t+\cc6,\cr
&\phi=-(\cc1t+\cc2)u+{y\over\beta}\left(\cc1x+\fft1\right)
+{\cc3x\over2\beta}+f_2(t),\cr}$$
if $\alpha =2\beta$, where $\cc1,\cc2,\ldots,\cc6$ are arbitrary
constants and
$f_1(t)$ and $f_2(t)$ are arbitrary differentiable functions
\cite{\refCMd}.

\subsection{Nonclassical Reductions}
Applying the nonclassical method to (\eqswwIIi) yields an additional
reduction in the case when $\alpha=\beta$, when it reduces to
(\eqblmp),
This nonclassical reduction is given by
$$u(x,y,t)= v(\eta,\tau)+w(\zeta,\tau)-\,{f(t)g^2(y)\over 2\beta}{\d
f\over\d t},\qquad
\eta=x+f(t)g(y),\quad \zeta=x-f(t)g(y),\quad \tau=t,\eqn{eqswwIIisr}$$
where $V(\eta,\tau)=v_{\eta}$ and $W(\zeta,\tau)=w_{\zeta}$ satisfy the
variable-coefficient KdV equations
$$\eqalignno{
&V_{\eta\eta\eta}-2\beta VV_{\eta}- V_\tau-{F(\tau)}V
+{\eta\over2\beta}\left\{{\d F\over\d \tau}+2F^2(\tau)\right\}
+Q(\tau)=0,&\eqnn{swwIIincsol}\cr
&W_{\zeta\zeta\zeta}-2\beta WW_{\zeta}- W_{\tau}
-{F(\tau)}W+{\zeta\over2\beta}
\left\{{\d F\over\d \tau}+2F^2(\tau)\right\}+Q(\tau)=0,
&\eqnn{swwIIiincsol}\cr}$$
with $F(\tau) = \bdot{f}(t)/f(t)$ and $Q(\tau)$ an arbitrary function.
It is
easily shown that (\swwIIincsol) and (\swwIIiincsol) satisfy the
necessary
conditions of the \p\ test due to Weiss \etal \cite{\refWTC} to be
completely
integrable and both equations can be transformed into the usual KdV
equation
(\eqkdv). In particular, if in (\swwIIincsol) and (\swwIIiincsol) we
set
$f(t)=\cc{}$, a constant, then we obtain solutions to (\eqswwIIi) with
$\alpha=\beta$ in terms of sums of solutions of the KdV equation
(\eqkdv), with
arguments $x\pm g(y)$, where $g(y)$ is an arbitrary function; this is
perhaps
the simplest family of solutions to (\eqblmp) found using this
nonclassical
reduction.

Applying the nonclassical method to (\eqswwIIii) yields an additional
reduction in the case when $\alpha=\beta$, given by
$$u(x,y,t)= v(\eta,\tau)+w(\zeta,\tau),\qquad
\eta=x+\kappa y,\qquad \zeta=x-\kappa y,\qquad
\tau=t,\eqn{eqswwIIiisr}$$
where $V(\eta,\tau)=v_{\eta}$ and $W(\zeta,\tau)=w_{\zeta}$ satisfy
$$\eqalignno{
&\kappa V_{\eta\eta\eta}-\beta\kappa VV_{\eta}-
V_\tau+Q(\tau)=0,&\eqnn{swwIIincsol}\cr &\beta\kappa
W_{\zeta\zeta\zeta}-\beta\kappa WW_{\zeta}-
W_{\tau}+Q(\tau)=0,&\eqnn{swwIIiincsol}\cr}$$
with $Q(\tau)$ an arbitrary function.

\section{Discussion}
In this paper we have discussed symmetry reductions and exact solutions
for the
shallow water wave equation (\eqgsww). In particular, for the special
case of
(\eqgsww) given by the {\hs} equation (\eqswwi), using the nonclassical
symmetry reduction method originally proposed by Bluman and Cole
\cite{\refBCa}, we obtained a family of solutions (\swwsol) which have
a rich
variety of qualitative behaviours. This is due to the freedom in the
choice of
the arbitrary function $f(t)$. One can choose $f_1(t)$ and $f_2(t)$
such
$|f_1(t)-f_2(t)|$ is exponentially small as $t\to-\infty$, yet $f_1(t)$
and
$f_2(t)$ are quite different as $t\to-\infty$, so that as $t\to-\infty$
the two
solutions are essentially the same, yet as $t\to\infty$ they are
radically
different. In Figure 1 we show that by a judicious choice of $f(t)$ we
can
exhibit a plethora of different solutions. We believe that these
results suggest
that solving the {\hs} equation (\eqswwi) numerically for initial
conditions
such as those in the solutions plotted in Figure 1 could pose some
fundamental
difficulties. An exponentially small change in the initial data yields
a
fundamentally different solution as $t\to\infty$. How can any numerical
scheme
in current use cope with such behaviour?

The solution (\swwsol) appears to be a nonlinear superposition of
solutions
suggesting that the {\hs} equation (\eqswwi) may be linearisable
through a
transformation to a linear \pde, analogous to the linearisation of
Burgers'
equation
$$ u_t = u_{xx} + 2uu_x, \eqn{burger} $$ which is mapped to the linear
heat
equation through the Cole-Hopf transformation \cite{\refCole,\refHopf}.
If so
then the solution (\swwsol) could be viewed as an artefact of the fact
that the
{\hs} equation (\eqswwi) is linearisable. However the {\hs} equation
(\eqswwi)
can be expressed as the compatibility condition of the third order
spectral
problem (\scpia,\scpib). Further the associated scattering problem
(\scpia) is
very similar to that for the Boussinesq equation which has been
thoroughly
studied by Deift \etal \cite{\refDTT}. This strongly suggests the {\hs}
equation
(\eqswwi) is solvable by inverse scattering. The spatial part of the
inverse
scattering formalism (\scpia) only defines $u$ up to an arbitrary
additive
function of $t$; this arbitrary function may be incorporated into $u$
using the
variable-coefficient transformation (\eqGalTr). Indeed the initial
value problem for the {\hs} equation (\eqswwi) is \underbar{\it not}
well-posed without the imposition of an additional constraint since if
$u(x,t)$
is a solution of (\eqswwi) satisfying the inital condition
$u(x,0)=\phi(x)$,
then so is
$$\~u(\~x,\~t)=u(x,t)+[g(t)-t]/\beta,\qquad \~x=x,\qquad \~t=
g(t),\eqn{eqGalTrr}$$ where $g(t)$ is \underbar{\it any} differentiable
function
such that $g(0)=0$. It appears likely that the inverse scattering
formalism for
the {\hs} equation (\eqswwi) will require that $u(x,t)$ satisfies a
constraint
of the form
$$\int_{-\infty}^{\infty}|u(x,t)|\,\d x < \infty,\eqn{eqcons}$$
for all $t$. It is well-known that such constraints are required in the
inverse
scattering formalism for $2+1$-dimensional equations such as the
Kadomtsev-Petviashvili equation (cf., \cite{\refVA}, see also
\cite{\refAC}
and the references therein).

Since the shallow water wave equation (\eqgsww) is invariant under
the variable-coefficient transformation (\eqGalTr) for all nonzero
$\alpha$ and $\beta$, one can take any solution of (\eqgsww) and using
(\eqGalTr) generate some interesting solutions. For example, one can
apply the
transformation (\eqGalTr) to the two-soliton solution (\swwiisol) and
thus
generate some exotic solutions for the \akns\ equation (\eqswwii)
analogous to
the solution (\swwsol) of the {\hs} equation (\eqswwi). Consequently
the above
remarks about difficulties in solving (\eqswwi) numerically apply to
(\eqgsww)
in general. Additionally, the inverse scattering formalism for
the {\akns} equation (\eqswwi) will probably require that $u(x,t)$
satisfies a
constraint such as (\eqcons).

Recently Ablowitz, Schober and Herbst \cite{\refASH} have shown that
the
focusing nonlinear \sch\ equation
$$\i u_t + u_{xx} + |u|^2u=0,\eqn{eqnls}$$ exhibits numerical chaos
created by
small errors on the order of roundoff. The results of Ablowitz \etal
\cite{\refASH} together with those given in this paper suggest that
numerical
analysts need to take care to ensure the accuracy of their programs.
Numerical
predictions of chaos may not always be what they seem!

\ack
It is a pleasure to thank Mark Ablowitz, Chris Cosgrove, Jim Curry,
Pilar
Est\'evez, Willy Hereman and Colin Rogers for several illuminating
discussions.
We also thank the Program in Applied Mathematics, University of
Colorado
at Boulder and the Department of Mathematics and Statistics, University
of
Pittsburgh for their hospitality during our visits whilst some of this
work was
done. The support of EPSRC (grant GR/H39420) is gratefully
acknowledged.

\def\refpp#1#2#3{\frenchspacing\nrm#1, #3 (#2).}
\def\refjl#1#2#3#4#5{\frenchspacing\nrm#1, {\frenchspacing\nit#3},\
{\nbf#4}, #5\ (#2).}
\def\refbk#1#2#3#4{\frenchspacing\nrm#1, ``{\nsl#3}'', #4 (#2).}
\def\refcf#1#2#3#4#5#6{\frenchspacing\nrm#1, in ``{\nsl#3}''
[\nrm#4],\ #5,\
#6 (#2).}

\def\fit{\frenchspacing\nit}
\def\refn#1{\item{\hbox to 20pt{\nrm\hfill[\expandafter \csname
#1\endcsname]\ }}}

\references

{\baselineskip=10pt\parindent=25pt

\refn{refAC}
\refbk{M.J. Ablowitz and P.A. Clarkson}{1991}{Solitons, Nonlinear
Evolution
Equations and Inverse Scattering}{{\fit L.M.S. Lect. Notes
Math.\/}, {\nbf 149}, C.U.P., Cambridge}
\refn{refAKNS}
\refjl{M.J. Ablowitz, D.J. Kaup, A.C. Newell and H. Segur}{1974}{Stud.
Appl.
Math.}{53}{249--315}
\refn{refARSa}
\refjl{M.J. Ablowitz, A. Ramani and H. Segur}{1978}{\PRL}{23}{333--338}
\refn{refARSb}
\refjl{M.J. Ablowitz, A. Ramani and H. Segur}{1980}{\jmp}{21}{715--721}
\refn{refASH}
\refjl{M.J. Ablowitz, C. Schober and B.M.
Herbst}{1993}{\PRL}{71}{2683--2686}
\refn{refVA}
\refjl{M.J. Ablowitz and J. Villarroel}{1991}{\sam}{85}{195--213}
\refn{refAndIb}
\refbk{R.L. Anderson and N.H. Ibragimov}{1979}{Lie-B\"acklund
Transformations in Applications}{SIAM, Philadelphia}
\refn{refBBM}
\refjl{T.B. Benjamin, J.L. Bona and J. Mahoney}{1972}{Phil. Trans. R.
Soc.
Lond. Ser. A}{272}{47--78}
\refn{refBCa}
\refjl{G.W. Bluman and J.D. Cole}{1969}{J. Math. Mech.}{18}{1025--1042}
\refn{refBK}
\refbk{G.W. Bluman and S. Kumei}{1989}{Symmetries and Differential
Equations}{{\fit Appl. Math. Sci.\/} {\nbf 81}, Springer-Verlag,
Berlin}
\refn{refBogi}
\refjl{O.I. Bogoyaviemskii}{1990}{Math. USSR Izves.}{34}{245--259}
\refn{refBogii}
\refjl{O.I. Bogoyaviemskii}{1990}{Russ. Math. Surv.}{45}{1--86}
\refn{refBLMP}
\refjl{M. Boiti, J.J-P. Leon, M. Manna and F.
Pempinelli}{1986}{\IP}{2}{271--279}
\refn{refBuchi}
\refcf{B. Buchberger}{1988}{Mathematical Aspects of Scientific
Software}{Ed.\
J.\ Rice}{Springer Verlag}{pp59--87}
\refn{refCHW}
\refjl{B. Champagne, W. Hereman and P. Winternitz}{1991}{Comp. Phys.
Comm.}{66}{319--340}
\refn{refPACrev}
\refpp{P.A. Clarkson}{1994}{``Nonclassical symmetry reductions for the
Boussinesq equation'', {\fit Chaos, Solitons \&\ Fractals\/}, to
appear}
\refn{refCK}
\refjl{P.A. Clarkson and M.D. Kruskal}{1989}{\jmp}{30}{2201--2213}
\refn{refCMa}
\refjl{P.A. Clarkson and E.L. Mansfield}{1994}{Physica D}{70}{250--288}
\refn{refCMc}
\refjl{P.A. Clarkson and E.L.
Mansfield}{1994}{Nonlinearity}{7}{975--1000}
\refn{refCMb}
\refpp{P.A. Clarkson and E.L. Mansfield}{1994}{``Algorithms for the
nonclassical method of symmetry reductions'', {\fit SIAM J. Appl.
Math.\/}, to
appear}
\refn{refCMd}
\refpp{P.A. Clarkson and E.L. Mansfield}{1994}{``Exact  solutions for
some
2+1-dimensional shallow water wave equations'', preprint, Department of
Mathematics, University of Exeter}
\refn{refCole}
\refjl{J.D. Cole}{1951}{Quart. Appl. Math.}{9}{225--236}
\refn{refCM}
\refjl{R. Conte and M. Musette}{1991}{\jmp}{32}{1450--1457}
\refn{refDTT}
\refjl{P. Deift, C. Tomei and E. Trubowitz}{1982}{Commun. Pure Appl.
Math.}{35}{567--628}
\refn{refDGRW}
\refjl{B. Dorizzi, B. Grammaticos, A. Ramani and P.
Winternitz}{1986}{\JMP}{27}{2848--2852}
\refn{refEF}
\refjl{A. Espinosa and J. Fujioka}{1994}{\JPSJ}{63}{1289--1294}
\refn{refGGKM}
\refjl{C.S. Gardner, J.M. Greene, M.D. Kruskal and R. Miura
M}{1967}{Phys. Rev. Lett}{19}{1095--1097}
\refn{refGNW}
\refjl{C.R. Gilson, J.J.C. Nimmo and R.
Willox}{1993}{\pl}{180A}{337--345}
\refn{refFush}
\refjl{W.I. Fushchich}{1991}{Ukrain. Math. J.}{43}{1456--1470}
\refn{refHere}
\refjl{W. Hereman}{1994}{Euromath Bull.}{1 {\nrm no. 2}}{45--79}
\refn{refHiet}
\refcf{J. Hietarinta}{1990}{Partially Integrable Evolution Equations in
Physics}{Eds. R. Conte and N. Boccara}{{\nit NATO ASI Series C:
Mathematical and Physical Sciences\/}, {\nbf 310}, Kluwer,
Dordrecht}{pp459--478}
\refn{refHirt}
\refcf{R. Hirota}{1980}{Solitons}{Eds. R.K. Bullough and P.J.
Caudrey}{{\nit Topics in Current Physics\/}, {\nbf 17},
Springer-Verlag, Berlin}{pp157--176}
\refn{refHI}
\refjl{R. Hirota and M. Ito}{1983}{\JPSJ}{52}{744--748}
\refn{refHS}
\refjl{R. Hirota and J. Satsuma}{1976}{J. Phys. Soc.
Japan}{40}{611--612}
\refn{refHopf}
\refjl{E. Hopf}{1950}{Commun. Pure Appl. Math.}{3}{201--250}
\refn{refInce}
\refbk{E.L. Ince}{1956}{Ordinary Differential Equations}{Dover, New
York}
\refn{refJM}
\refjl{M. Jimbo and T. Miwa}{1983}{Publ. R.I.M.S.}{19}{943--1001}
\refn{refLU}
\refjl{S.B. Leble and N.V. Ustinov}{1994}{\IP}{210}{617--633}
\refn{refLW}
\refjl{D. Levi and P. Winternitz}{1989}{\jpa}{22}{2915--2924}
\refn{refMD}
\refpp{E.L. Mansfield}{1993}{``{\ntt diffgrob2}: A symbolic algebra
package
for analysing systems of PDE using Maple", {\ntt ftp
euclid.exeter.ac.uk},
login: anonymous, password: your email address, directory: {\ntt
pub/liz}}
\refn{refMF}
\refpp{E.L. Mansfield and E.D.\ Fackerell}{1992}{``Differential
Gr\"obner
Bases",  preprint {\nbf 92/108}, Macquarie University, Sydney,
Australia}
\refn{refMcLO}
\refjl{J.B. McLeod and P.J. Olver}{1983}{SIAM J. Math.
Anal.}{14}{488--506}
\refn{refMLD}
\refjl{M. Musette, F. Lambert and J.C.
Decuyper}{1987}{\JPA}{20}{6223--6235}
\refn{refOlver}
\refbk{P.J. Olver}{1993}{Applications of Lie Groups to Differential
Equations}{Second Edition, {\fit Graduate Texts  Math.} {\nbf
107}, Springer-Verlag, New York}
\refn{refORi}
\refjl{P.J. Olver and P. Rosenau}{1986}{\pl}{114A}{107--112}
\refn{refORii}
\refjl{P.J. Olver and P. Rosenau}{1987}{SIAM J. Appl.
Math.}{47}{263--275}
\refn{refPer}
\refjl{H. Peregrine}{1966}{J. Fluid Mech.}{25}{321--330}
\refn{refReida}
\refjl{G.J. Reid}{1990}{\jpa}{23}{L853--L859}
\refn{refReidb}
\refjl{G.J. Reid}{1991}{Europ. J. Appl. Math.}{2}{293--318}
\refn{refRW}
\refpp{G.J. Reid and A. Wittkopf}{1993}{``A Differential Algebra
Package
for Maple'', {\ntt ftp 137.82.36.21} login: anonymous, password: your
email address, directory: {\ntt pub/standardform}}
\refn{refSchw}
\refjl{F. Schwarz}{1992}{Computing}{49}{95--115}
\refn{refTP}
\refjl{K.M. Tamizhmani and P. Punithavathi}{1990}{\JPSJ}{59}{843--847}
\refn{refTop}
\refjl{V.L. Topunov}{1989}{Acta Appl. Math.}{16}{191--206}
\refn{refWeiss}
\refjl{J. Weiss}{1983}{\jmp}{24}{1405--1413}
\refn{refWTC}
\refjl{J. Weiss, M. Tabor and G. Carnevale}{1983}{\jmp}{24}{522--526}
\refn{refWW}
\refbk{E.E. Whittaker and G.M. Watson}{1927}{Modern Analysis}{Fourth
Edition,
C.U.P., Cambridge}
\refn{refWint}
\refcf{P. Winternitz}{1993}{Integrable Systems, Quantum Groups, and
Quantum Field
Theories}{Eds. L.A. Ibort and M.A. Rodriguez}{{\nit NATO ASI Series C},
{\nbf 409}, Kluwer, Dordrecht}{pp425--495}

\vfill\eject

\figures

\noindent{\bf Figure 1.}\quad
\hangindent=16pt\hangafter=1
The solution (\swwsoli) where
\hfill\break{\hbox to 30pt{(i),\hfill}}$f(t)=\tfr12t$,
\hfill\break{\hbox to
30pt{(ii),\hfill}}$f(t)=\tfr12t-\exp\left(\tfr15t-1\right)$,
\hfill\break{\hbox to 30pt{(iii),\hfill}}$f(t)=\tfr12t+[1-\tanh t]\sin
t$,
\hfill\break{\hbox to
30pt{(iv),\hfill}}$\tfr12t-\exp\left(-t^2\!/100\right)\sin t$,
\hfill\break{\hbox to 30pt{(v),\hfill}}$f(t)=\tfr12t+6\pi^{-1}\tan^{-1}
t$,
\hfill\break{\hbox to 30pt{(vi),\hfill}}$f(t)=\tfr12\sqrt{t^2+20}$,
\hfill\break{\hbox to 30pt{(vii),\hfill}}$f(t)=f(t)=\tfr14t(1-\tanh
t)+\tfr12[2\sin(\tfr23 t)+3](1+\tanh t)$,
\hfill\break{\hbox to 30pt{(viii),\hfill}}$f(t)=\tfr14t(1-\tanh t)$.

\smallskip\noindent{\bf Figure 2.}(a)\quad
\hangindent=16pt\hangafter=1
The solution (3.12) of the \hs\ equation (1.8) and its $x$-derivative
for
$c=3$.
\smallskip\noindent{\bf Figure 2.}(b)\quad
\hangindent=16pt\hangafter=1
The $x$-derivative of solution (4.6) of the \hs\ equation (1.8) for
$\kappa_1=2$,
$\kappa_2=1.7$ and $\kappa_1=\tfr34$, $\kappa_2=\tfr23$.

\smallskip\noindent{\bf Figure 3.}\quad
\hangindent=16pt\hangafter=1
The solution (3.17) of equations (3.13) and its $x$-derivative for
$c=\tfr14$.

\smallskip\noindent{\bf Figure 4.}\quad
\hangindent=16pt\hangafter=1
The solution (3.18) of equations (3.14) and its $x$-derivative for
$c=\tfr14$.

\smallskip\noindent{\bf Figure 5.}\quad
\hangindent=16pt\hangafter=1
The solution (4.6) of the \hs\ equation (1.8).
\hfill\break{\hbox to 30pt{(i),\hfill}}$\kappa_1=2$, $\kappa_2=1.7$
\hfill\break{\hbox to 30pt{(ii),\hfill}}$\kappa_1=\tfr34$,
$\kappa_2=\tfr23$,
\hfill\break{\hbox to 30pt{(iii),\hfill}$\kappa_1=1$,
$\kappa_2=(11+3\sqrt{93}\,)/20$ ($A_{12}=0$),
\hfill\break{\hbox to 30pt{(iv),\hfill}}$\kappa_1=0.8$,
$\kappa_2=\tfr15(2+3\sqrt{7}\,)$ ($A_{12}=0$).

\smallskip\noindent{\bf Figure 6.}\quad
\hangindent=16pt\hangafter=1
The solution (4.9) of the \hs\ equation (1.8).

\smallskip\noindent{\bf Figure 7.}\quad
\hangindent=16pt\hangafter=1
Two soliton solutions (4.10) of the \akns\ equation (1.7).

\vfill\eject

\bye